\documentclass[twocolumn]{aastex63}

\pdfoutput=1

\newcommand{\ChaNGa}{\textsc{ChaNGa}~} 

\begin{document}

\title{Quenching timescales of dwarf satellites around Milky Way-mass hosts}
\shortauthors{Akins et al.}
\shorttitle{Quenching Timescales of Dwarf Satellites}

\correspondingauthor{Hollis B. Akins} 
\email{hollis.akins@gmail.com}

\author[0000-0003-3596-8794]{Hollis B. Akins}
\affiliation{Department of Physics, Grinnell College, 1116 Eighth Ave., Grinnell, IA 50112, USA}

\author[0000-0001-6779-3429]{Charlotte R. Christensen}
\affiliation{Department of Physics, Grinnell College, 1116 Eighth Ave., Grinnell, IA 50112, USA}

\author[0000-0002-0372-3736]{Alyson M. Brooks}
\affil{Department of Physics \& Astronomy, Rutgers, the State University of New Jersey, 136 Frelinghuysen Rd., Piscataway, NJ 08854, USA}

\author[0000-0002-9581-0297]{Ferah Munshi}
\affil{Department of Physics \& Astronomy, The University of Oklahoma, 440 W. Brooks St., Norman, OK 73019, USA}

\author[0000-0001-8301-6152]{Elaad Applebaum}
\affil{Department of Physics \& Astronomy, Rutgers, the State University of New Jersey, 136 Frelinghuysen Rd., Piscataway, NJ 08854, USA} 

\author{Anna Engelhardt}
\affiliation{Department of Physics, Grinnell College, 1116 Eighth Ave., Grinnell, IA 50112, USA}

\author{Lucas Chamberland}
\affiliation{Department of Physics, Grinnell College, 1116 Eighth Ave., Grinnell, IA 50112, USA}

\received{August 7, 2020}
\revised{January 31, 2021}
\accepted{February 1, 2021}
\submitjournal{ApJ}

\begin{abstract} 
	Observations of the low-mass satellites in the Local Group have shown high fractions of gas-poor, quiescent galaxies relative to isolated dwarfs, implying that the host halo environment plays an important role in the quenching of dwarf galaxies. 
	In this work, we present measurements of the quenched fractions and quenching timescales of dwarf satellite galaxies in the DC Justice League suite of 4 high-resolution cosmological zoom-in simulations of Milky Way-mass halos. 
	We show that these simulations accurately reproduce the satellite luminosity functions of observed nearby galaxies, as well as the variation in satellite quenched fractions from $M_* \sim 10^5~\mathrm{M}_\sun$ to $10^{10}~\mathrm{M}_{\sun}$. 
	We then trace the histories of satellite galaxies back to $z \sim 15$, and find that many satellites with $M_* \sim 10^{6-8}~\mathrm{M}_\sun$ quench within $\sim 2$ Gyr of infall into the host halo, while others in the same mass range remain star-forming for as long as $5$ Gyr. 
	We show that this scatter can be explained by the satellite's gas mass and the ram pressure it feels at infall. 
	Finally, we identify a characteristic stellar mass scale of ${10}^{8}~\mathrm{M}_{\sun}$ above which infalling satellites are largely resistant to rapid environmental quenching. 
\end{abstract}


\section{Introduction}\label{sec:intro}

The shallow potentials of dwarf galaxies make them uniquely sensitive laboratories for understanding the physics of galaxy formation. 
In particular, their sensitivity to their environment makes dwarfs useful in studying the quenching of star formation.
Observations show that nearly all Local Group (LG) satellites with $M_* < 10^8~\mathrm{M}_\sun$ are quenched \citep{wetzel_rapid_2015,weisz_star_2015}, while isolated dwarfs are not \citep{geha_stellar_2012}.  
This dichotomy is supported by observations of the gas content of LG dwarfs, with atomic hydrogen (HI) undetected in all satellites within $270~\mathrm{kpc}$ of their host \citep{grcevich_local_2009, spekkens_dearth_2014}.
While quenching may act differently outside of the LG \citep[see][]{geha_saga_2017}, these observations firmly establish that the halo environment of a massive LG galaxy can greatly alter the star-formation of its dwarf satellites.

The rapid removal of cold disk gas by ram pressure has traditionally been seen as the dominant quenching mechanism, and has been shown to quench infalling satellites in many simulations \citep[e.g.][]{murakami_babul_1999,mayer_simultaneous_2006, slater_mass_2014, bahe_star_2015, kazantzidis_effects_2017, simpson_quenching_2018}. 
Stellar feedback may contribute to this process, both by increasing the efficiency of stripping by expelling and heating central gas \citep{bahe_star_2015, kazantzidis_effects_2017} and by reducing the central satellite mass and therefore reducing its potential \citep{zolotov_baryons_2012}.  
Tidal stripping provides another mechanism by which the satellite potential may be reduced, in addition to driving gas loss \citep{mayer_simultaneous_2006}.
Additionally, at higher satellite masses, the suppressed gas accretion from cosmological inflows due to the presence of a massive halo may be the dominant mechanism quenching satellites \citep{mcgee_overconsumption_2014, wheeler_surprising_2014}, though on longer timescales. 
While the effects of these individual processes are well documented, it remains unclear how they conspire to quench dwarf satellites across different mass scales and environments.  

A fruitful metric for disentangling the roles of these various quenching processes is the timescale on which quenching occurs.
Therefore, many authors have estimated the quenching timescales necessary to reproduce the high quenched fractions observed in the LG.
The simplest statistical models assume that quenching occurs some ``delay time'' after the satellite's accretion into the virialized volume of host halo. 
For low-mass satellites ($M_* < 10^8~\mathrm{M}_\sun$), the LG quenched fractions are high ($\gtrsim 80\%$), and can only be reproduced if these delay times are small, on the order of $1$--$2$ Gyr \citep{slater_mass_2014,fillingham_taking_2015, wetzel_rapid_2015}.
Expanding these models, \citet{weisz_star_2015} have used \textit{HST} imaging to derive star formation histories (SFHs) and quenching times for 38 LG dwarfs. 
Combining these results with estimates of infall times determined via abundance-matching with simulations, the authors found that many low-mass LG satellites quench prior to infall. 
More recently, \citet{fillingham_characterizing_2019} have combined these SFHs with infall times determined from \textit{Gaia} proper motions, and found results for low-mass satellites ($10^5 < M_*/\mathrm{M}_\sun < 10^8$) generally consistent with the $1$--$2$ Gyr timescales inferred from the quenched fractions. 

The rapid, $1$--$2$ Gyr quenching timescales for low-mass satellites suggest the removal of satellite gas by ram-pressure and tidal stripping, rather than simply by gas consumption in the absence of accretion.
In the latter case, quenching is expected to occur on the cold gas depletion timescale, which is typically much longer than $\sim 2$ Gyr for star-forming dwarf galaxies \citep{huang_gas_2012, fillingham_taking_2015}.
The rapid quenching timescales for low-mass satellites have been reproduced in the Auriga suite of cosmological simulations, and ram pressure appears to be the dominant quenching mechanism in these cases \citep{simpson_quenching_2018}. 
However, ram pressure alone may not be able to quench satellites on these rapid timescales \citep{emerick_gas_2016}, and may require the aid of stellar feedback, outflows, and continued gas consumption due to star formation. 
For more massive satellites, in which ram pressure is likely inefficient at removing gas, observational estimates suggest much longer quenching timescales ($\gtrsim 8~\mathrm{Gyr}$), more consistent with the gas depletion timescales expected for quenching driven primarily by a lack of accretion \citep{wetzel_galaxy_2013, wheeler_surprising_2014, phillips_mass_2015}.
However, even at these higher masses, satellite quenching may still be affected by ram-pressure stripping \citep{bahe_star_2015} and stellar feedback-driven outflows \citep{mcgee_overconsumption_2014}.
Therefore, while the quenching timescales for LG satellites are a key constraint on quenching processes, determining the relative roles of different quenching processes requires combining timescale data with further analysis of the satellites.

At present, most studies of the quenching timescale are observational in nature, and they use simulations and abundance-matching techniques only to constrain the infall times of LG dwarfs \citep[e.g.][]{wheeler_surprising_2014, weisz_star_2015, fillingham_characterizing_2019}.
Such observations are limited by uncertainties in quenching and infall times, the number of observable satellites, and an inability to observe the galaxy at the time of infall.
Full hydrodynamic simulations can aid in the interpretation of observations by allowing for more precise measurements of quenching and infall times and the direct measurement of various satellite properties during their accretion, such as 
ram-pressure stripping \citep{simpson_quenching_2018} and morphological transformation \citep{kazantzidis_effects_2017}.
Such work can also be combined with a broader analysis of satellite properties in simulated galaxies, such as their SFHs \citep{Wetzel2016, garrison-kimmel_star_2019, digby_star_2019}, radial distribution \citep{garrison-kimmel_2017, Richings_2020, samuel_profile_2020}, velocity alignment \citep{Riley_2019}, velocity dispersion \citep{Buck_2019}, and kinematics \citep{brooks_why_2014} to produce a more complete picture of host-satellite interaction.
Together, these results paint a picture in which interactions between the satellite and the host galaxy's circumgalactic medium and gravitational potential are responsible for not only shaping the satellite's SFH but also their structure, even to the point of disruption.
Nevertheless, many of the simulations that have been used to analyze satellite quenching achieve high resolution only by focusing on individual satellites \citep[e.g.][]{mayer_simultaneous_2006, emerick_gas_2016, kazantzidis_effects_2017}, and others that simulate larger cosmological volumes must do so at the cost of resolution \citep[e.g.][]{bahe_star_2015}. 

\begin{deluxetable*}{lCCCCCCCC}
\tablecaption{Host Halo Properties at $z=0$\label{tab:simulations}}
\tablehead{
	\colhead{Simulation} & \colhead{$M_{\mathrm{vir}}~(\mathrm{M}_\sun)$} & \colhead{$R_{\mathrm{vir}}~(\mathrm{kpc})$} & \colhead{$M_{\mathrm{HI}}~(\mathrm{M}_\sun)$} &  \colhead{$M_R$ (AB mag)} & \colhead{$M_{*,R}~(\mathrm{M}_\sun)$\tablenotemark{a}} & \colhead{$M_{*,\text{sim}}~(\mathrm{M}_\sun)$\tablenotemark{b}} &  \colhead{sSFR (yr$^{-1}$)\tablenotemark{c}}}
\startdata
Sandra & 2.7 \times 10^{12}  & 432 & 1.4 \times 10^{10}   & -23.1 & 1.2 \times 10^{11} & 1.9 \times 10^{11}  & 8.3\times10^{-11}\\
Ruth   & 1.3 \times 10^{12}  & 340 & 1.9 \times 10^{10}   & -22.2 & 5.3 \times 10^{10} & 1.0\times 10^{11}   & 5.3\times10^{-11}\\
Sonia  & 1.2 \times 10^{12}  & 333 & 2.2 \times 10^{10}  & -21.9 & 3.8 \times 10^{10} & 9.0\times 10^{10}   & 3.1\times10^{-11}\\
Elena  & 8.3 \times 10^{11}  & 293 & 4.6 \times 10^{8}  & -22.2 & 5.2 \times 10^{10} & 9.0 \times 10^{10}  & 6.4\times10^{-11}
\enddata	
\tablenotetext{a}{Stellar mass derived from the Johnson-Cousins $R$-band assuming a mass-to-light ratio of 1.}
\tablenotetext{b}{Stellar mass calculated directly from the simulation.}
\tablenotetext{c}{Average rate of star formation over the past 100 Myr divided by the actual stellar mass of the galaxy.}
\end{deluxetable*}

Simulating satellites within the larger host environment is critical for facilitating comparisons to observations, and high resolution is important for modeling quenching across the full range of dwarf satellite masses. 
Therefore, we analyze here the DC Justice League suite of simulations: four cosmological zoom-in simulations of Milky Way-mass galaxies, run at sufficiently high resolution that all classical dwarf galaxies are resolved with at least 50 star particles. 
This ``near-mint" resolution is comparable to that of the NIHAO ultra high-resolution simulations \citep{Buck_2019}, and the central halos contain over a million dark matter particles and between 34 and 94 million total particles. 
\citet{simpson_quenching_2018} similarly measured the quenching timescales of dwarf satellites around Milky Way-mass galaxies.
They found a strong trend in the quenched fraction versus stellar mass, with a transition threshold between primarily quenched to primarily star-forming of $M_* \sim 10^8~\mathrm{M}_\sun$.
Below this mass threshold, most galaxies quenched on timescales $\lesssim 1$ Gyr and ram-pressure stripping was ubiquitous.
While \citet{simpson_quenching_2018} includes a larger sample of host galaxies than that presented here, we provide higher resolution (we note, however, that \citet{simpson_quenching_2018} does show convergence of the quenched fraction vs. stellar mass to resolutions similar to ours.).
Unlike the smoothed particle hydrodynamic code here, that paper uses a moving-mesh code, AREPO \citep{springel_arepo_2010}, designed to model shocks and hydrodynamic instabilities with high fidelity.
We also model stellar feedback through the locally-dependent subgrid blastwave prescription as opposed to a phenomenological wind model, and we allow for the natural formation of a multiphase ISM, rather than employ a two-phase subgrid model for the ISM.
The latter is particularly important for reproducing the resistance of the cold molecular ISM to ram-pressure stripping \citep{Tonnesen2009}.

In this paper, we compare the satellite quenched fraction as a function of stellar mass to observations of the LG and beyond, and we determine the quenching timescales for individual dwarf satellites as a function of their mass.
By analyzing the properties of the satellites at their time of accretion in relation to their quenching timescales, we infer the mass regimes over which different quenching processes likely dominate.
The structure of the paper is as follows. 
In \S\ref{sec:methods} we present the suite of simulations. 
In \S\ref{sec:comparisons} we validate the simulations with comparisons to observations and discuss differences. 
In \S\ref{sec:timescales} we present the quenching and infall times of our dwarf satellites and discuss the roles of different quenching processes. We summarize our conclusions in \S\ref{sec:conclusion}.

\begin{figure*}
	\includegraphics[width=\linewidth]{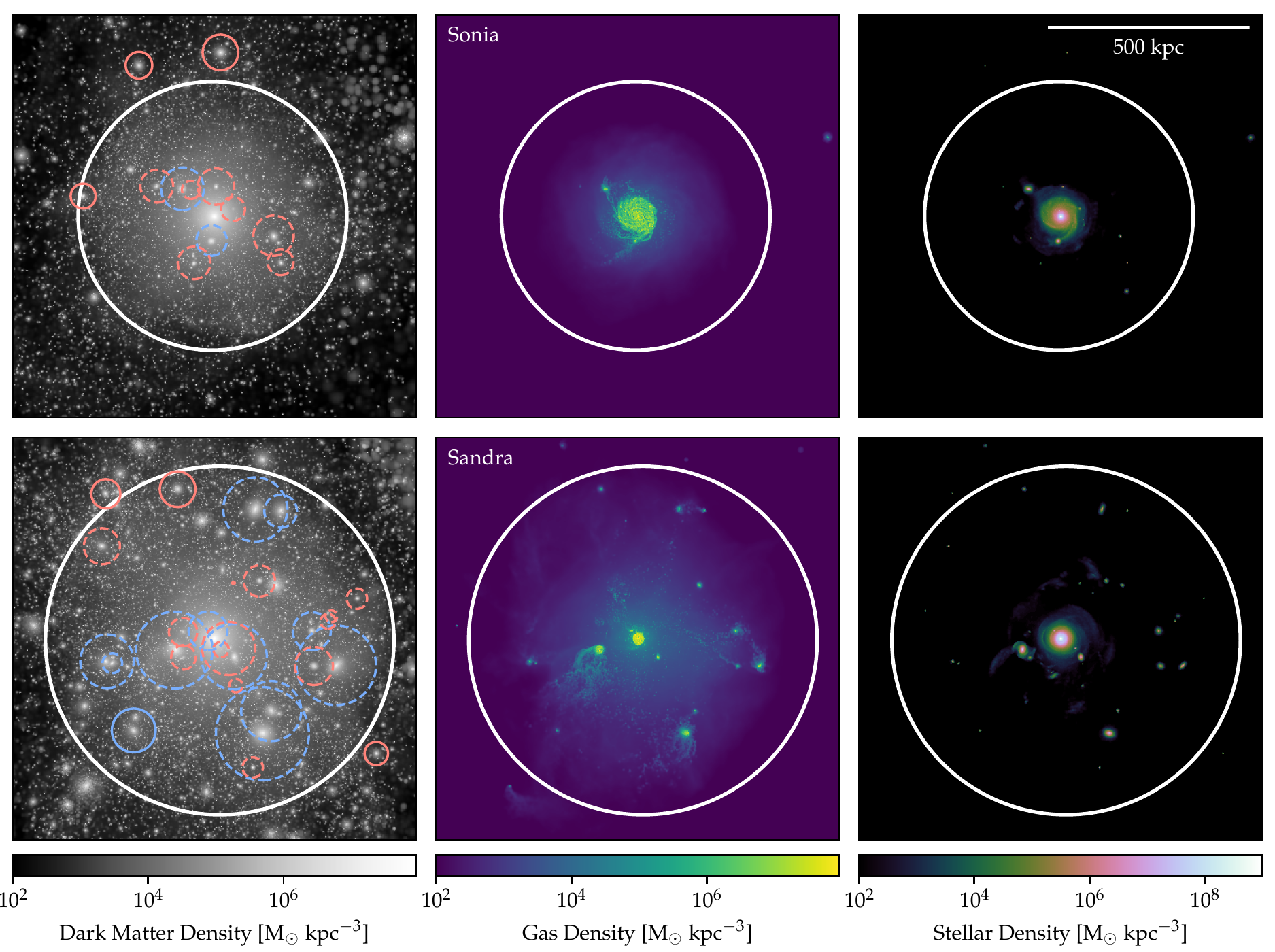}
	\caption{Density plots of dark matter (left), gas (middle), and stars (right) in Sonia (top) and Sandra (bottom) at $z=0$. 
	In each panel, the central disk is aligned face-on and the densities are averaged along the line-of-sight.
	Dark matter halos are outlined in the left panel, and the central halo is outlined for both gas and stars.  
	Satellite (central) galaxies are shown as dashed (solid) lines, and quenched (star-forming) galaxies are shown in red (blue). 
	Several dwarf satellites can be seen, many of which are gas-poor and quenched.}\label{fig:density}	
\end{figure*} 

\section{Methods}\label{sec:methods}
 
We conduct our analysis on the DC Justice League suite of 4 ``near-mint''-resolution, cosmological zoom-in, smooth particle hydrodynamics (SPH) simulations of Milky Way-mass disk galaxies and their surroundings. 
The properties of the host halos in the Justice League simulations are shown in Table~\ref{tab:simulations}. 
These simulations were previously introduced in \citet{bellovary_multimessenger_2019}. 
Here we discuss the simulation code and the post-processing analysis conducted on the simulation output. 

\subsection{Simulations}

All simulations were generated using the tree+SPH code \ChaNGa \citep{menon_adaptive_2015}, which scales efficiently up to 100,000 cores. 
\ChaNGa is the successor to the N-body gravity-tree code PKDGRAV \citep{stadel_cosmological_2001} and SPH code \textsc{gasoline} \citep{wadsley_gasoline_2004,wadsley_gasoline2_2017}. 
\ChaNGa models Kelvin-Helmholtz instabilities in shearing flows by using the geometric mean density in the SPH force expression \citep{ritchie_multiphase_2001, Keller2014}.
This method generally minimises numerical surface tension associated with density discontinuities, including those found in Kelvin-Helmholtz instabilities.
Correctly modeling the instabilities and shocks of the satellite halo gas as it passes through the host galaxy CGM is key to correctly modeling the gas loss rates due to ram pressure stripping \citep{Quilis2000}.
\ChaNGa also allows for thermal diffusion across gas particles with a thermal diffusion coefficient of 0.03 \citep{shen_enrichment_2010}. 

The simulations were integrated from $z=149$ to $z=0$ in a fully cosmological context assuming Planck 2015 cosmological parameters \citep[$\Omega_0=0.3086$, $\Omega_b = 0.04860$, $\Lambda = 0.6914$, $h=0.67$, $\sigma_8=0.77$;][]{planck_2016}.
In order to achieve high resolution while maintaining the effects of the large-scale environment, the initial conditions were generated using the ``zoom-in'' volume renormalization technique \citep{katz_hierarchical_1993}.
The four main halos were selected from $50^3~\mathrm{Mpc}^3$ dark matter-only volumes to be a representative sample of  the range  of masses, halo spins and local densities considered to be representative of our Milky Way local environment.  
Halos that had major mergers  at $z<0.5$  or  were  in the infall zone of group and clusters were avoided and no requirement for a nearby similar mass  companion was imposed. 
The highest-resolution region extends to $\sim 1.5$ times the $z=0$ virial radius of the main halo.
The simulations have a force softening resolution of $170$ pc. 
Dark matter particle masses are $4.2\times 10^4~\mathrm{M}_\sun$, gas particle masses are $2.7\times 10^4~\mathrm{M}_\sun$, and star particles form with masses of $8000~\text{M}_\sun$. 

\ChaNGa follows non-equilibrium abundances of H (including H$_2$) and He species. 
The integration of these chemical species and the associated heating and cooling is described in \citet{christensen_implementing_2012} and references therein.
Photoionization and heating rates are implemented using a uniform, time-dependent cosmic UV background adapted from \citet{haardt_radiative_2012}. 
In this model, cosmological HII regions overlap at $z \sim 6.7$ (13 Gyr ago), indicating the completion of reionization.
We note that this model is known to ionize and heat the intergalactic medium too early, which primarily affects gas thermodynamics for $z > 6$ \citep{Onorbe2017}. 
Cooling beyond that from hydrogen and helium is provided by metal lines assuming photoionization equilibrium \citep{shen_enrichment_2010}.
Oxygen and iron are tracked independently, and metals are diffused across particles based on a subgrid turbulent mixing model \citep{shen_enrichment_2010} with a metal diffusion constant of 0.03.

Star particles represent simple stellar populations with a \citet{kroupa_initial_2002} initial mass function.
    Star formation in \ChaNGa is implemented probabilistically according to local H$_2$ abundance, gas density, and gas temperature, as described in \citet{christensen_implementing_2012}.
Briefly, the star formation efficiency factor is given by $c^* =  0.1 f_{\mathrm{H}_2} = 0.1 \frac{X_{\mathrm{H}_2}}{X_{\mathrm{H}_2} + X_{\mathrm{HI}}}$, where $X_{\mathrm{HI}}$ and $X_{\mathrm{H}_2}$ are the mass fraction of atomic and molecular hydrogen, respectively.
Star formation is only allowed for particles with temperatures $< 10^3$ K and densities $> 0.1$ amu cm$^{-3}$, although these constraints are largely superseded by the dependency of $c^*$ on H$_2$, and most stars form at densities $> 100$ amu cm$^{-3}$.

Energy from Type II supernovae (SNe II) is distributed as thermal energy to surrounding gas particles according to the ``blastwave'' model \citep{stinson_star_2006}, assuming $\eta_{\mathrm{SN}} = 1.5\times10^{51}$ ergs released per SN.  
Cooling of the affected gas particles is temporarily disabled to match the theoretical timescale of the ``snowplow'' phase of the supernova \citep{mckee_theory_1977}, avoiding the rapid radiative cooling that would render feedback negligible.
This SNe feedback model provides the entirety of stellar feedback as radiative pressure \citep[e.g.][]{stinson_magicc_2012, hopkins_galaxies_2014} and other forms of early stellar feedback are not separately included.
This stellar feedback model relies entirely on the local properties of the gas, rather than the overall halo potential, but still produces mass-loading factors consistent with energy-driven winds \citep{christensen_innout_2016}.
It has been critical for reproducing the observed properties of dwarf galaxies, including cored dark matter profiles \citep{governato_bulgeless_2010}, and reconciling the Too Big To Fail problem \citep{brooks_why_2014}.
Energy and metals from SNe Ia are also deposited in gas particles within a smoothing kernel, but cooling is not disabled.
Stellar winds also return mass, at the metallicity of the star particles, to gas particles within the smoothing kernel assuming the mass loss rates from \citep{Weidemann1987}.

This version of \ChaNGa also includes supermassive black hole formation \citep{bellovary_first_2011}, growth, mergers, and feedback \citep{tremmel_romulus_2017}. 
However, none of the black holes accrete at high enough rates during their history for black holes to play a critical role in the quenching of star formation \citep{bellovary_multimessenger_2019}.

\subsection{Post-processing Analysis}

In order to select individual galaxies from simulation snapshots, we use \textsc{amiga's halo finder} \citep{knollmann_ahf_2009}, which identifies regions of over-density and assigns halo ownership to gravitationally bound particles. 
We compute the virial radius ($R_{\mathrm{vir}}$) of each halo as the radius at which the enclosed density drops below 200 times the background density, $\rho_b$ = $\Omega_M \rho_{\mathrm{crit}}$.
Galaxy properties are calculated from all particles within the virial radius, excluding subhalos. 
Table~\ref{tab:simulations} shows basic $z=0$ properties of the host galaxies in the 4 Justice League simulations, which are named Sandra, Ruth, Sonia, and Elena. 
All four galaxies have disk morphology at z = 0 but have different masses and merger histories.
Most dramatically, Elena experienced a merger at $z \sim 0.5$, which led to its low $z = 0$ HI mass, although it retains a low-surface brightness disk.
Figure~\ref{fig:density} shows line-of-sight averaged density plots of the dark matter, gas, and stars in Sandra and Sonia.

For each Justice League simulation, we use the database-generation software \textsc{tangos} \citep{pontzen_tangos_2018} to track particles across snapshots and generate merger trees for all halos containing stars at $z=0$. 
For each halo, we determine the major progenitor, defined as the halo which contains the majority of dark matter particles from the corresponding halo in the subsequent snapshot, back to $z \approx 15$. 
We limit the merger tree to progenitor halos with  $> 1000$ dark matter particles, corresponding to a halo mass of $\sim 10^7~\mathrm{M}_\sun$, approximately the resolution limit of our simulations. 
All further analysis is conducted in Python using the \textsc{pynbody} package \citep{pontzen_pynbody_2013}.

\section{Observational Comparisons}\label{sec:comparisons}

\subsection{Satellite Luminosity Functions}

\begin{figure}
	\includegraphics[width=\linewidth]{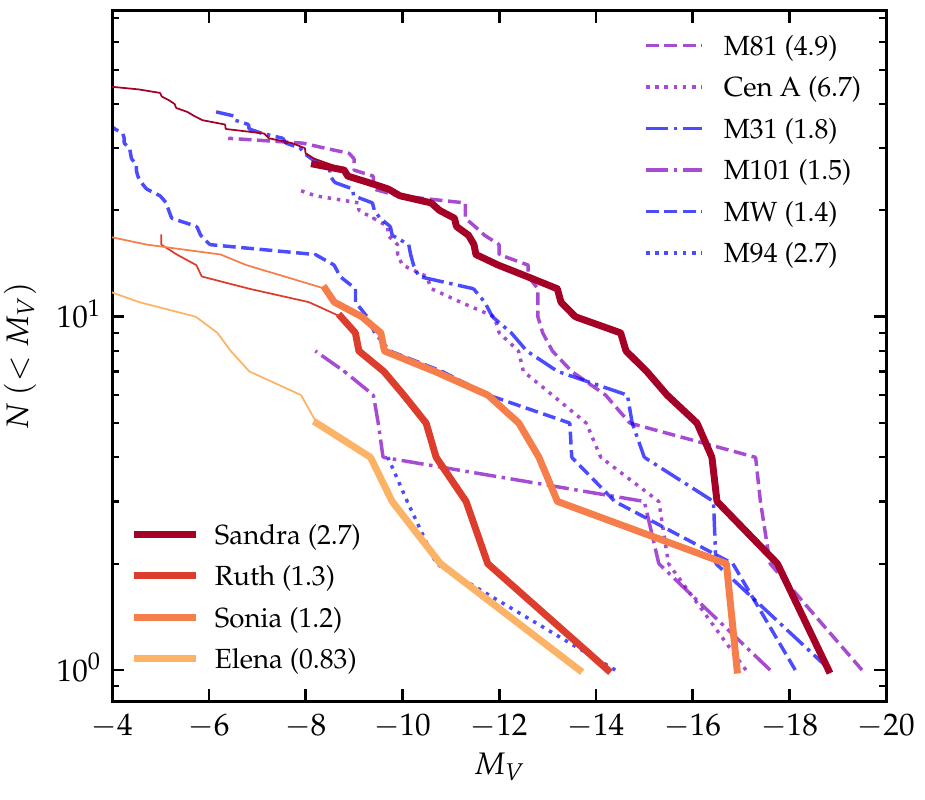}
	\caption{Cumulative satellite luminosity functions for the Justice League simulations. 
	Red/orange lines show simulated LFs, with thin lines extending below the high-resolution limit of 50 star particles. 
	Observational luminosity functions, presented for comparison, are shown with blue/purple and in dashed/dotted lines. 
	Observational data comes from the updated version of the \citet{mcconnachie_observed_2012} catalog for the MW and M31, \citet{crnojevic_faint_2019} for Centaurus A, and \citet{bennet_m101_2019} for M101. 	
	For M94, we include the two satellites recently discovered by \citet{smercina_lonely_2018}, along with two previously known satellites KK 160 and IC 3687 \citep{crnojevic_faint_2019}. 
	Data for M81 comes from \citet{chiboucas_confirmation_2013} and includes the recently discovered satellite from \citet{smercina_d1005_2017}, using the $M_{r'}$ to $M_V$ conversion provided by \citet{crnojevic_faint_2019}. 
	Values in parentheses are the host galaxy's virial (orbital) mass, in $10^{12}~\mathrm{M}_\sun$, from \citet{Karachentsev2014} for observations.}\label{fig:lumfunc}
\end{figure}

Figure~\ref{fig:lumfunc} shows the cumulative satellite luminosity functions (LFs) of our four Justice League simulations, alongside observed LFs for nearby host-satellite systems.
While we show the LFs down to $M_V \sim -4$, we focus our analysis on satellites above our resolution limit, which encompasses the ``classical dwarfs'' ($M_V$ brighter than $-8$). 
LFs derived from the recently-completed ``mint''-resolution runs of these same simulations, with roughly eight-times-higher mass resolution, match these down to $M_V \sim -7$ \citep{applebaum_ultrafaint_2020}. 
To aid in the comparison to observations, we list the stellar masses of the simulated galaxies as they would be calculated from $M_R$ assuming a mass-to-light ratio of 1.
The actual stellar masses of the simulated galaxies may be found in Table~\ref{tab:simulations}.

Although the LFs of our simulations vary widely (from $\sim 5$ to $\sim 25$ satellites brighter than $M_V \sim -8$), they are consistent with the range of observed LFs for MW-analog hosts. 
The LFs of our simulated sample are roughly ordered by host halo mass, with the more massive halos generally hosting more satellites with a magnitude of $M_V \sim -8$ or brighter, as expected \citep{trentham_dwarf_2009, nickerson_luminosity_2013, carlsten2020luminosity}. 
Additionally, our simulations are not susceptible to the missing satellite problem \citep{klypin_where_1999}, as we do not see systematic overproduction of luminous satellites. 
While each galaxy hosts on the order of 10-50 dark matter subhalos with $M_{\mathrm{vir}} > 10^8~\mathrm{M}_\sun$, the occupation fraction for the least massive halos ($M_{\mathrm{vir}} < 10^8~\mathrm{M}_\sun$) is low, with $< 5\%$ of halos hosting $> 10$ star particles. 
This low occupation fraction is a result of the inclusion of baryonic physics in the simulation, which has been shown to reduce the number of luminous satellites \citep{brooks_baryonic_2013} by core-creation and/or tidal stripping.

The most populous system, Sandra, is consistent with the LFs of the most massive galaxies: the star-forming disk galaxies M31 \citep{mcconnachie_observed_2012} and M81 \citep{chiboucas_confirmation_2013}, and the elliptical radio galaxy, Centaurus A \citep{crnojevic_faint_2019}.
As a star-forming disk galaxy with a strong central bar, Sandra is more similar in morphology to M31 and M81 than Centaurus A.
Its $2.7 \times 10^{12}~\mathrm{M}_\sun$ virial mass is also closer to the orbital masses of M31 and M81 ($1.76 \times 10^{12}$ and $4.89 \times 10^{12}~\mathrm{M}_\sun$, respectively) than that of Centaurus A \citep[$6.71 \times 10^{12}~\mathrm{M}_\sun$;][]{Karachentsev2014}.
The moderate mass hosts, Ruth and Sonia, have LFs most consistent with the MW \citep{mcconnachie_observed_2012} and M101 \citep{bennet_m101_2019}, and their spiral morphologies and virial masses are likewise similar to those of the MW \citep[$M_{\mathrm{orbit}} = 1.35 \times 10^{12}~\mathrm{M}_\sun$;][]{Karachentsev2014} and M101 \citep[$M_{\mathrm{orbit}}=1.47 \times 10^{12}~\mathrm{M}_\sun$;][]{Karachentsev2014}.
The least massive host, Elena, has a LF consistent with the unique and sparsely populated M94 system \citep{smercina_lonely_2018}. 
However, it differs both in its virial mass, which is less than a third of the $2.67 \times 10^{12}~\mathrm{M}_\sun$ orbital mass measured for M94 \citep{Karachentsev2014}, and its morphology.
Unlike the other hosts in our sample, Elena is undergoing quenching due to a merger at $z \sim 0.5$ and now contains an extended low surface brightness disk.
Therefore, a larger sample of lower stellar mass satellite hosts may be necessary to identify a better analogue for Elena.

\subsection{Quenched Fractions}

To further ensure that our star-formation model accurately reproduces observed quenching in dwarf galaxies, we also measure the mass trend in the fraction of quenched satellites.
Here, and throughout this paper, we define a ``quenched'' galaxy as one with a specific star-formation rate ($\mathrm{sSFR} = \mathrm{SFR}/M_*$) of $< 10^{-11}~\mathrm{yr}^{-1}$, where we define the SFR as the average rate of star formation within the Amiga-identified halo over the past 100 Myr. 
Analysis of the star formation histories indicates that while it is possible for satellites to drop below this threshold only temporarily, in almost all cases it indicates a genuine period of quiescence with $\mathrm{SFR} = 0~\mathrm{M}_\sun\,\mathrm{yr}^{-1}$ for $\gtrsim$ 1 Gyr. 
This choice of threshold has been adopted by many other studies using simulations \citep[e.g.][]{bahe_star_2015, pallero_tracing_2019}, and some observational studies use this threshold with spectroscopically derived measures of the sSFR \citep[e.g.][]{wetzel_galaxy_2013}. 
Other observational studies define quenching directly from spectroscopic parameters, specifically H$\alpha$ emission and the D$_n4000$ index \citep[e.g.][]{geha_stellar_2012, geha_saga_2017}. 
Alternatively, spectroscopic information may be used to estimate the atomic hydrogen richness $M_{\mathrm{HI}}/M_*$ in dwarfs, a useful indicator of star formation \citep[e.g.][]{grcevich_local_2009, spekkens_dearth_2014, wetzel_rapid_2015}.
To facilitate observational comparisons, we computed the best-fit power-law relationship between the sSFR and atomic hydrogen richness in our simulations and found that our sSFR threshold corresponds to $M_{\mathrm{HI}}/M_* \approx 0.2$.

\begin{figure}
	\includegraphics[width=\linewidth]{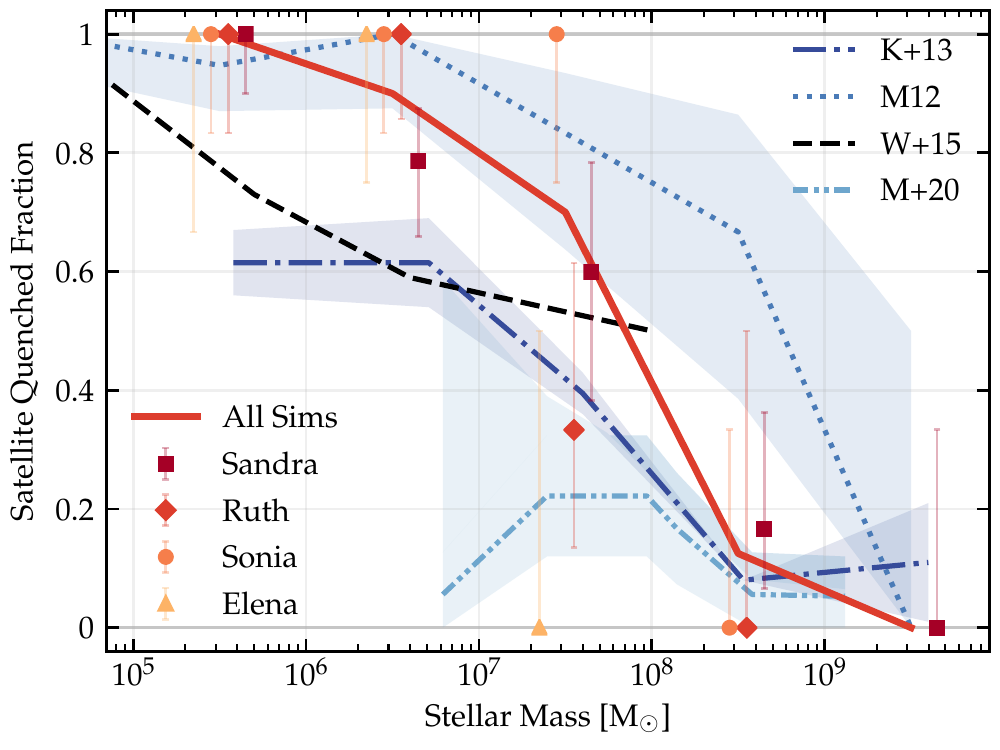} 
	\caption{Quenched fractions for Justice League satellites in 1 dex bins of stellar mass. Points are offset from the center of the bin to make them distinguishable. 
	Individual simulations are indicated with points of varying shape, while the solid line represents the total for all four simulations. 
	All simulated data are limited to galaxies with more than 50 star particles, and error bars represent 68\% uncertainty in the binomial proportion via the Wilson Score Interval \citep{Wilson1927}. 
	Observational data from \citet[][M+20]{mao_saga_2020},
	\citet[][M12]{mcconnachie_observed_2012}, \citet[][W+15]{weisz_star_2015}, and \citet[][K+13, as compiled by Weisz]{karachentsev_updated_2013} are presented for comparison. 
	Shaded regions represent the given uncertainty in observed quenched fractions. 
	or the \citet{mao_saga_2020} data, this includes both shot noise and an incompleteness correction.
	Our data generally agree with observations of the LG and other nearby galaxies, and a relationship between quenched fraction and  galaxy mass is apparent.
	}\label{fig:comparisons}	
\end{figure}

Figure~\ref{fig:comparisons} shows the fraction of quenched satellites in each of the Justice League simulations, in 1 dex bins of stellar mass.  
In order to avoid sample contamination by galaxies far from the central host, we restrict our study to galaxies within the virial radius of their host halo. 
While we note that the virial radii of our hosts are generally larger than the $300$ kpc estimate typically used for the MW, our results are not sensitive to this choice as we find few satellites at large radial distances.
Actual satellite stellar masses from the simulations are reported on this plot.

Observational data from \citet{mcconnachie_observed_2012}, \citet{karachentsev_updated_2013}, \citet{weisz_star_2015}, and \citet{mao_saga_2020} are shown for comparison.
These studies cover different samples of satellite populations: while \cite{mcconnachie_observed_2012} and \citet{weisz_star_2015} sample LG satellites, the Nearby Galaxy Catalog \citep{karachentsev_updated_2013} includes satellites of hosts out to 11 Mpc and the Satellites Around Galactic Analogs (SAGA) survey \citep{geha_saga_2017,mao_saga_2020} includes satellites of hosts between 20 and 40 Mpc away. 
For the \citet{mcconnachie_observed_2012} LG sample, we define the quenched fraction as the fraction of satellites within 300 kpc of their host and with $M_{\mathrm{HI}}/M_* < 0.2$, the HI threshold that best corresponds to our sSFR threshold.  
We note that this threshold produces identical quenched fractions as the threshold of $0.1$ adopted by \citet{fillingham_taking_2015} and \citet{wetzel_rapid_2015}. 
\citet{weisz_star_2015} also use a 300 kpc radius to classify satellites in the LG, and use the absence of detected HI to define a satellite as quenched.
Data from the Nearby Galaxy catalog \citep{karachentsev_updated_2013} is also compiled in \citet{weisz_star_2015}.
Quenching for this sample is based on galaxy morphological type $T$ \citep[a numerical code according to classification by][]{devaucouleurs_third_1991}.
Galaxies with $T < 0$ (largely dSph) are classified as quenched, whereas galaxies with $T \geqslant 0$ are not. 
For some dwarfs with transitioning morphological types (dTrans), this definition becomes ambiguous. 
As such, the upper/lower bounds for the \citet{karachentsev_updated_2013} line in Figure~\ref{fig:comparisons} represent quenched fractions with all/none of the dTrans galaxies classified as quenched.
Finally, \citet{mao_saga_2020} define a satellite as quenched if its spectrum shows no H$\alpha$ emission. 
We have verified using the radiative transfer code \textsc{Sunrise} \citep{jonsson_highresolution_2010} that, in all cases but one, our star forming galaxies above the SAGA completeness limit of $M_r < -12.3$ also have detectable H$\alpha$ emission and our quenched galaxies do not.

Figure~\ref{fig:comparisons} shows generally good agreement between quenched fractions in the Justice League simulations and nearby galaxy observations. 
Our simulations fall within the range of observations, fitting the data from \citet{mcconnachie_observed_2012} well at low masses and \citet{weisz_star_2015} and \citet{karachentsev_updated_2013} better at higher masses. 
Given the widely varying samples of satellites in these observations, it is startling that our results agree as well as they do, and suggest that morphology and HI content are well correlated with star-formation rates. 
Our results are also remarkably consistent with the simulations presented in \citet{simpson_quenching_2018}, despite substantial differences in physical prescriptions used.
Specifically, allowing the formation of a multi-phase ISM using non-equilibrium chemistry models, as we do here, does not appear to have significantly changed the quenched fractions compared to their use of a sub-grid two-phase, pressure-equilibrium model for the ISM.

Figure~\ref{fig:comparisons} also highlights variation between the individual Justice League simulations. 
The quenched fractions of satellites around the individual hosts vary widely, particularly at stellar masses of $\sim 10^{7-8}~\mathrm{M}_\sun$. 
This halo-to-halo scatter is consistent with observed variations across the MW and M31, where we see drastic differences in quenched fractions. 
At $M_* \sim 10^{8-9}~\mathrm{M}_\sun$, all M31 satellites (M32 \& NGC 205) are quenched, while all MW satellites (the Magellanic clouds) are star-forming \citep{mcconnachie_observed_2012}.

\subsection{Comparisons to the SAGA Survey}\label{sec:SAGA}

While the quenched fractions of satellites of different stellar masses in our simulations are consistent with observations of the LG and other nearby galaxies, there is tension with the results of the SAGA survey \citep{geha_saga_2017, mao_saga_2020}.
Although individual simulated halos such as Elena (and, potentially, Ruth) are consistent with the quenched fractions of the SAGA survey, our sample as a whole contains more quenched galaxies with $M_* \sim 10^{7-8}~\mathrm{M}_\sun$ than SAGA thus far (Figure~\ref{fig:comparisons}).
While a full analysis must wait until the complete SAGA results, we discuss possible explanations for the differing quenched fractions here.

In order to produce a more detailed and fair comparison to SAGA, we estimate SDSS $r$-band magnitudes for the simulated satellites using {\sc PARSEC} \citep{Bressan2012} isochrones and calculate the H$\alpha$ emission (the star formation tracer SAGA uses) with the radiative transfer code \textsc{Sunrise} \citep[though we caution that different assumptions about stellar population and radiative transfer models may produce slightly different results]{jonsson_highresolution_2010}. 
SAGA has a completeness limit of $M_r = -12.3$, which corresponds to  $M_* \sim 10^6~\mathrm{M}_\Sun$ for star-forming galaxies and $\sim 10^7~\mathrm{M}_\Sun$ for quenched galaxies.
Of the 123 satellites observed by SAGA brighter than this limit, 105 are actively star-forming. 
Our simulated sample includes 26 satellites with $M_r < -12.3$, 19 of which are star-forming and 18 of which are predicted to have H$\alpha$ emission.
This difference in quenched fraction is concentrated at the faint-end, and our brightest quenched satellite has an $M_r = -14.5$.

Differences between SAGA hosts, the LG, and the host galaxies in our sample may explain some or all of this discrepancy in the satellite quenched fraction.
First, host virial masses are difficult to constrain observationally, and it is possible that the SAGA sample may emphasize a lower range of halo masses than the LG or our simulated sample.
Lower-mass host galaxies are known to have lower fractions of quenched satellites \citep{wetzel_galaxy_2012, phillips_mass_2015, bahe_star_2015}, which may help explain the lack of quenching among SAGA satellites.
Indeed, our least-massive Milky Way analog, Elena ($M_{\mathrm{vir}} = 8.5 \times 10^{11}~\mathrm{M}_\Sun$), has no quenched satellites with $M_* \sim 10^7$ --$10^8~\mathrm{M}_\Sun$, similar to what is observed for SAGA galaxies.
In the particular case of Elena, the lack of quenched satellites is driven largely by the low total number of satellites and the lower ram pressure they experience from the low-density CGM, both a product of the lower halo mass.

Second, the host halo's environment may affect its fraction of quenched satellites through group preprocessing \citep{wetzel_satellite_2015}, which may be more prevalent around higher-mass hosts \citep{jung_origin_2018}. 
The Milky Way-Andromeda system is somewhat uniquely situated as a close pair.
In contrast, while not explicitly disbarring a Milky Way-Andromeda like system, the selection criteria for SAGA hosts emphasizes relative isolation in order to reduce uncertainty in identifying satellites.
Specifically, no galaxy brighter than $M_K$ + 1 of the host galaxy magnitude could be within 1$^\circ$ of the host. Nor could a massive ($> 5 \times 10^{12}~\mathrm{M}_\sun$) galaxy be within two virial radii of the SAGA hosts.
Our simulated sample of host galaxies are more isolated than the LG: none are in a close pair and the distances to the closest galaxies with $M_{\mathrm{vir}} > 10^{11.5}~\mathrm{M}_\Sun$ are 1.68 Mpc (Sandra), 2.37 Mpc (Ruth), 2.55 Mpc (Sonia), and 6.14 Mpc (Elena).
While our higher-mass hosts may be in richer environments than the SAGA galaxies, more consistent with the LG, the particular isolation of Elena is more consistent with SAGA.
This is similar to the isolated, sparsely-populated systems M94 and M101, which have quenched fractions consistent with SAGA hosts \citep{bennet_m101_2019}. 
This result suggests a relationship between host halo mass, environment, number of satellites, and satellite quenched fraction that may help to explain the SAGA results. 

Finally, we stress that the completeness limit of the SAGA survey likely precludes detection of fainter quenched satellites. 
Our simulations, in agreement with data from the LG, suggest that satellites of MW-analog hosts transition from primarily star-forming to primarily quenched at lower masses. 
While the precise mass at which this transition occurs may be different in the SAGA hosts, it likely lies near the completeness limit for quenched galaxies. 
Furthermore, the satellite luminosity functions of the SAGA hosts generally show fewer faint ($-12.3 > M_r > -14$) satellites than recent observations of the Local Volume \citep{carlsten2020luminosity}. 
This, combined with the fact that the discrepancy between our results and those of SAGA is concentrated at $M_r > -14.5$, suggests that the SAGA results may be missing quenched satellites near their completeness limit.
Indeed, the incompleteness correction provided by \citet{mao_saga_2020} (shown in our Figure~\ref{fig:comparisons}), which assumes that all potential undiscovered satellites are quenched, is more inline with our simulations.
We predict that more complete observations in this transition range, and deeper observations fainter than $M_r = -12.3$, will uncover a trove of quenched satellites.
The complete SAGA survey of $\sim 100$ hosts will provide valuable observational data for understanding the efficiency of quenching outside of the LG.

\section{Quenching Timescales}\label{sec:timescales}

In this section, we present the relative times of quenching and infall for the Justice League satellites discussed above. 
The quenching timescale, commonly defined as the time a satellite remains star-forming after infall, measures the efficiency of environmental quenching and may hint at the underlying quenching processes.  
\subsection{Timescale Definitions}

We define the quenching time ($t_{\mathrm{quench}}$) as the lookback time at which a galaxy's sSFR last crossed below $10^{-11}~\mathrm{yr}^{-1}$. 
To determine $t_{\mathrm{quench}}$, we compute each galaxy's star formation history as the instantaneous sSFR of the main progenitor over time. 
Uncertainties in $t_{\mathrm{quench}}$ are computed from the quenching times with different choices of sSFR threshold, namely 0 yr$^{-1}$ and $2\times 10^{-11}~\mathrm{yr}^{-1}$. 
This is intended to flag those galaxies that undergo particularly slow quenching, for which the exact quenching epoch is difficult to constrain. 

Using a sSFR threshold to define quenching follows the methodology used in  \S\ref{sec:comparisons} to determine the fraction of satellites that are quenched, and we have demonstrated that this definition yields results in agreement with observations.
However, it is difficult to determine the historical values of sSFR through observations; instead, many observers define the quenching time as $t_{90}$, the lookback time at which a galaxy formed 90\% of its present-day stellar mass \citep[e.g.][]{weisz_star_2015, fillingham_characterizing_2019}.
As a check, we also compute $t_{90}$ for our satellites. 
Though not shown, adopting this definition does not qualitatively change our results, as the $t_{90}$ values are on average $1.32$ Gyr earlier than $t_{\mathrm{quench}}$.

We define the infall time ($t_{\mathrm{infall}}$) as the lookback time to the subhalo's first crossing of its host's virial radius ($R_{\mathrm{vir}}$). 
While many galaxies fall within 1 $R_{\mathrm{vir}}$ multiple times, we use the first crossing as the traditional picture of stripping-dominated quenching attributes the majority of gas removal to the first pericentric passage \citep{slater_confronting_2013}.
As with $t_{\mathrm{quench}}$, uncertainties are computed by varying the choice of threshold, namely from 0.9 to 1.1 $R_{\mathrm{vir}}$. 
For both infall and quenching times, these uncertainties are added in quadrature with the systematic uncertainty from the mean difference in snapshot times. 

\subsection{Results}

\begin{figure}
	\centering
	\includegraphics[width=\linewidth]{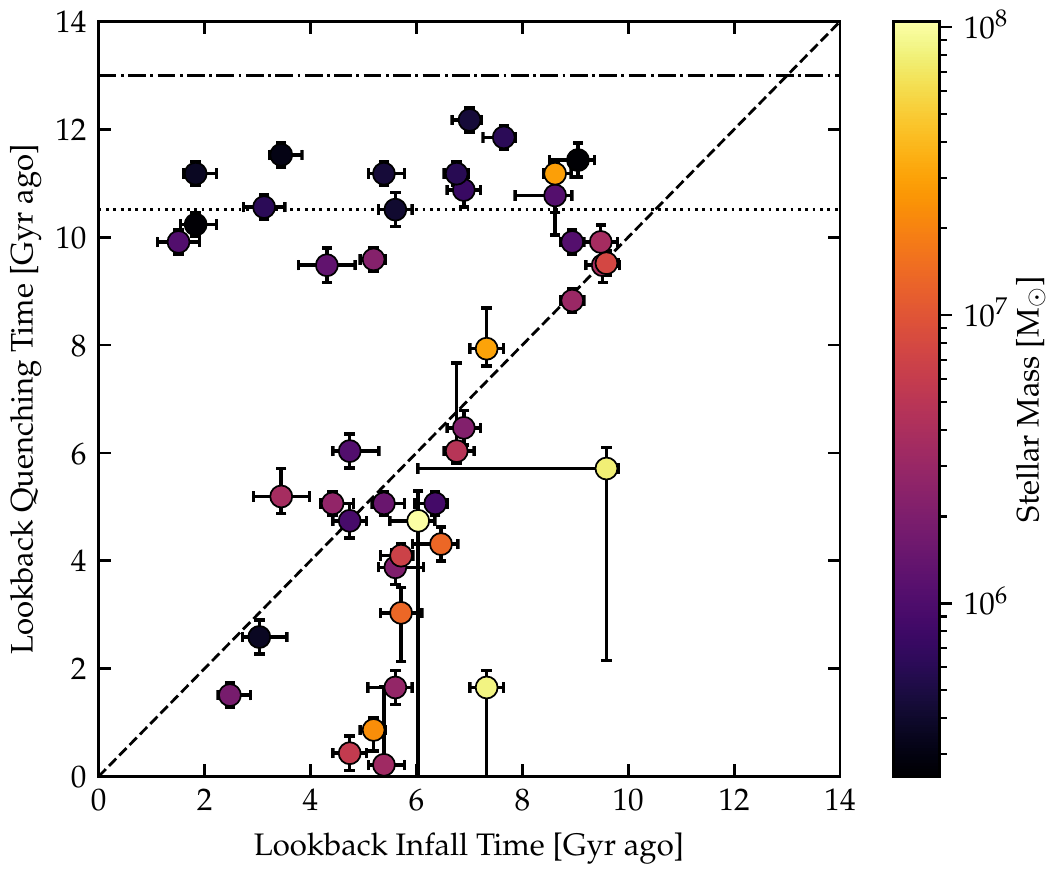}
	\caption{
		Lookback quenching time ($t_{\mathrm{quench}}$) vs.~lookback infall time ($t_{\mathrm{infall}}$) for all quenched galaxies surviving to $z=0$ above our resolution limit.
		Error bars indicate sensitivity to the choice of threshold; that is, upper (lower) limits for quenching time use $2\times 10^{-11}~\mathrm{yr}^{-1}$ ($0~\mathrm{yr}^{-1}$) rather than $10^{-11}~\mathrm{yr}^{-1}$, and upper (lower) limits for infall time use 1.1 (0.9) $R_{\mathrm{vir}}$ rather than 1 $R_{\mathrm{vir}}$. 
		Error bars also include systematic uncertainty from the mean difference between snapshot times. Points are colored by stellar mass. 
		The dashed line divides the plot into quenching before infall (left) and after infall (right). 
		The dot-dash line indicates the end of reionization, while the dotted line indicates the epoch of the peak UV heating rate ($z= 2$).
		Many satellites quench after infall, and those that quench before infall do so largely independent of their infall time. \\
	}\label{fig:QI}
\end{figure}

Figure~\ref{fig:QI} shows $t_{\mathrm{quench}}$ vs.~$t_{\mathrm{infall}}$ for all quenched satellites surviving to $z=0$ in the Justice League simulations. 
Galaxies falling to the right of the diagonal line on this plot were quenched after infall, while those to the left were quenched prior to infall.
We see that many of the lowest-mass galaxies quench early on ($t_{\mathrm{quench}} \gtrsim 9$ Gyr), but with little correlation to infall time, indicating that quenching occurred independently of the larger halo \citep{rodriguezwimberly_suppression_2019}. 
This may be a direct result of cosmic reionization \citep[as in][]{brown_quenching_2014}, though the time delay suggests a more indirect process. 
One possibility is that reionization suppresses gas accretion and the remaining gas is subsequently expelled by stellar feedback \citep{benitez-llambay_imprint_2015} or self-shielded and consumed in star-formation \citep{katz_how_2020}.
Another possibility is heating from the UV background, which peaked in our model around $z\sim 2$ \citep{haardt_radiative_2012}. 
Though not the primary subject of this work, these low-mass, early quenching galaxies are discussed in \citet{applebaum_ultrafaint_2020} and will be analyzed further in future work.
The lone high-mass ($M_* \sim 3\times10^7~\mathrm{M}_\sun$) galaxy that quenches this early experienced a period of intense starburst early in the Universe, reaching a peak SFR of $\sim 0.1~\mathrm{M}_\sun~\mathrm{yr}^{-1}$ and likely consuming the bulk of its gas. 
Importantly, we also see a population of predominantly higher-mass galaxies that quench after infall into the host halo (along and to the right of the diagonal line), to be discussed later.

Figure~\ref{fig:QI} also shows a few satellites that quench shortly before infall into their host halo (just to the left of the diagonal line and with $t_{\mathrm{quench}} \lesssim 8$ Gyr). 
The lowest-mass galaxy of these four ($M_* \sim 10^6~\mathrm{M}_\sun$) is a reionization fossil that experienced momentary reignition prior to infall, but all seem to feel the effects of environment beyond $1~R_{\mathrm{vir}}$. 
This extended environmental effect likely comes from some combination of group preprocessing \citep{wetzel_satellite_2015}, unique elliptical orbital trajectories \citep{simpson_quenching_2018}, and direct ram-pressure stripping from gas filaments \citep{bahe_why_2013}, and may be able to reignite quenched satellites in addition to quenching star-forming ones \citep{wright_reignition_2019}. 

\begin{figure}
\includegraphics[width=\linewidth]{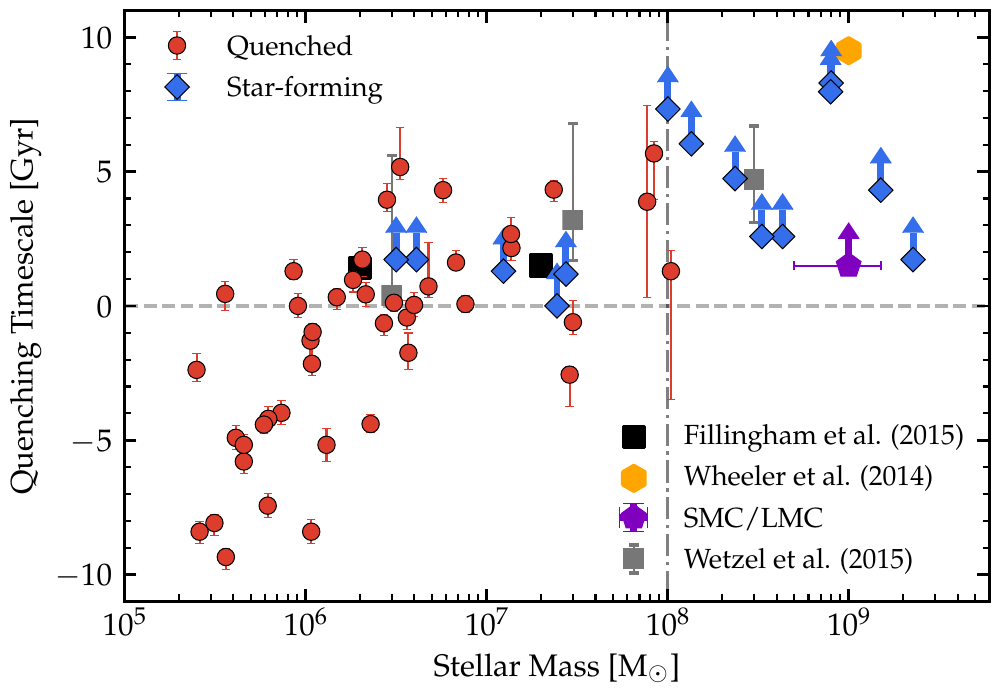}	
	\caption{
	Quenching timescale ($t_{\mathrm{infall}} - t_{\mathrm{quench}}$) vs.~stellar mass. 
	For galaxies that are quenched at $z=0$ (red circles), the timescale represents the time after infall until quenching, i.e. negative values indicate quenching prior to infall. 
	For galaxies that are star-forming at $z=0$ (blue diamonds), this is the lookback infall time and can be interpreted physically as a lower limit on the possible quenching timescale. 
	As in Figure~\ref{fig:QI}, error bars indicate threshold variation and systematic uncertainty (added in quadrature) and galaxies with less than 50 star particles are excluded. 
	Observational estimates from \citet{fillingham_taking_2015}, \citet{wetzel_rapid_2015}, and \citet{wheeler_surprising_2014} are shown in black squares, grey squares, and orange hexagons, respectively.  
	An estimate of the quenching timescale for the SMC/LMC system is shown in purple, using the infall time estimate from \citet{kallivayalil_thirdepoch_2013}. 
	A clear stellar mass threshold of $\sim 10^8~\mathrm{M}_\sun$ (indicated with a dot-dash line) is apparent, above which our satellites are largely resistant to rapid environmental quenching. 
	}\label{fig:timescales}
\end{figure}

While some galaxies certainly do quench prior to infall, higher-mass galaxies typically quench shortly after.
These quenching times correspond to a median satellite distance at $t_{\rm quench}$ of $0.6~R_{\rm vir}$, consistent with the stripping model of \citet{fillingham_taking_2015}.
This mass dependence is more readily apparent in Figure~\ref{fig:timescales}, which shows the quenching timescale $t_{\mathrm{infall}} - t_{\mathrm{quench}}$ vs.~ stellar mass. 
Galaxies which have become satellites but remain star-forming at $z=0$ are shown on this plot, as their lookback infall time can be interpreted as a lower limit on the potential quenching timescale. 
Many of the lowest-mass satellites ($M_* \lesssim 3\times 10^6~\mathrm{M}_\sun$) quench long before infall, likely due to cosmic reionization in the early universe \citep{brown_quenching_2014, weisz_star_2014, rodriguezwimberly_suppression_2019}.

A second stellar mass threshold of $\sim 10^8~\mathrm{M}_{\sun}$ is apparent in Figure~\ref{fig:timescales}, above which satellites can remain star-forming for as much as $\sim 7$ Gyr after infall. 
The implied lower limits on the quenching timescales are consistent with the unquenched Magellanic clouds, which experienced infall $\sim 1.5$ Gyr ago \citep[][their Figure 11]{kallivayalil_thirdepoch_2013}. 
Additionally, these lower limits are consistent with observations of higher-mass dwarfs from the NSA/SDSS catalog, which show long timescales and inefficient environmental quenching \citep{wheeler_surprising_2014}. 
It is true that satellites that remain star-forming after infall can continue to build up stellar mass and will consequently have higher $z=0$ stellar masses than those that quench. 
However, this stellar mass threshold appears to be primarily based on the properties of the galaxies at infall, rather than being a side-effect of star formation histories after infall.
As we show and discuss in \S\ref{sec:processes}, the galaxies that remain star-forming to $z=0$ typically have higher gas fractions and halo masses at infall.
Only two of our satellites with $M_*(z=0) > 10^8~\mathrm{M}_\sun$ have $M_*(t_\mathrm{infall}) < 10^8~\mathrm{M}_\sun$, and these are the two lowest-mass satellites above our $z=0$ threshold. 
The remaining high-mass satellites all form $\lesssim 40$\% of their $z=0$ stellar mass in the time since infall.

In contrast, for intermediate-mass satellites ($M_* \sim 10^{6-8}~\mathrm{M}_\sun$), we predominantly see rapid quenching timescales of $\lesssim 2$ Gyr. 
This is consistent with observational estimates of LG timescales from  \citet{wetzel_rapid_2015} and \citet{fillingham_taking_2015, fillingham_characterizing_2019} and the simulations of \citet{simpson_quenching_2018}.
\citet{wetzel_rapid_2015} and \citet{fillingham_taking_2015} use LG quenched fractions and semi-analytical models of quenching to infer the quenching times of LG satellites. 
They combine these inferred quenching times with infall times estimated from the ELVIS \citep{Garrison-Kimmel2014} simulations to determine the timescales reported on Figure~\ref{fig:timescales}. 
Our results broadly agree with the averages provided by \citet{wetzel_rapid_2015} and \citet{fillingham_taking_2015}.

Though the rapid quenching timescales we see in the Justice League simulations reaffirm LG observations, it is important to recognize the spread in the timescale at a given stellar mass. 
Figure~\ref{fig:timescales} also shows several intermediate-mass satellites that take more than $2$ Gyr to quench. 
These galaxies, despite being relatively low in mass ($M_* < 10^{8}$) are able to maintain high sSFRs for as much as $5$ Gyr after infall. 
These quenching timescales are generally within the uncertainty range provided by \citet{wetzel_rapid_2015}.  
We note that one of these satellites experiences some ongoing, but bursty star formation in the 4 Gyr since infall, and could be considered to have quenched earlier (see \S\ref{sec:processes}). 
The few unquenched intermediate-mass satellites, which have no LG analog, accreted recently ($\lesssim 1$ Gyr ago), and the lower limits on their quenching timescales are largely consistent with the timescales estimated for this mass range by \citet{fillingham_taking_2015}.

The  $M_* \sim 10^8~\mathrm{M}_{\sun}$ threshold seems to indicate a characteristic mass scale at which the efficiency of the dominant quenching process changes. 
As we discuss in the following section, some gas removal process (likely ram-pressure stripping) dominates the quenching of galaxies with $M_* \sim 10^{6-8}~\mathrm{M}_\sun$.
For higher mass galaxies, we argue that this process becomes much less efficient at removing gas, allowing for extended star formation following the satellite's accretion.

\subsection{Quenching Processes}\label{sec:processes}

\begin{figure*}
\includegraphics[width=\linewidth]{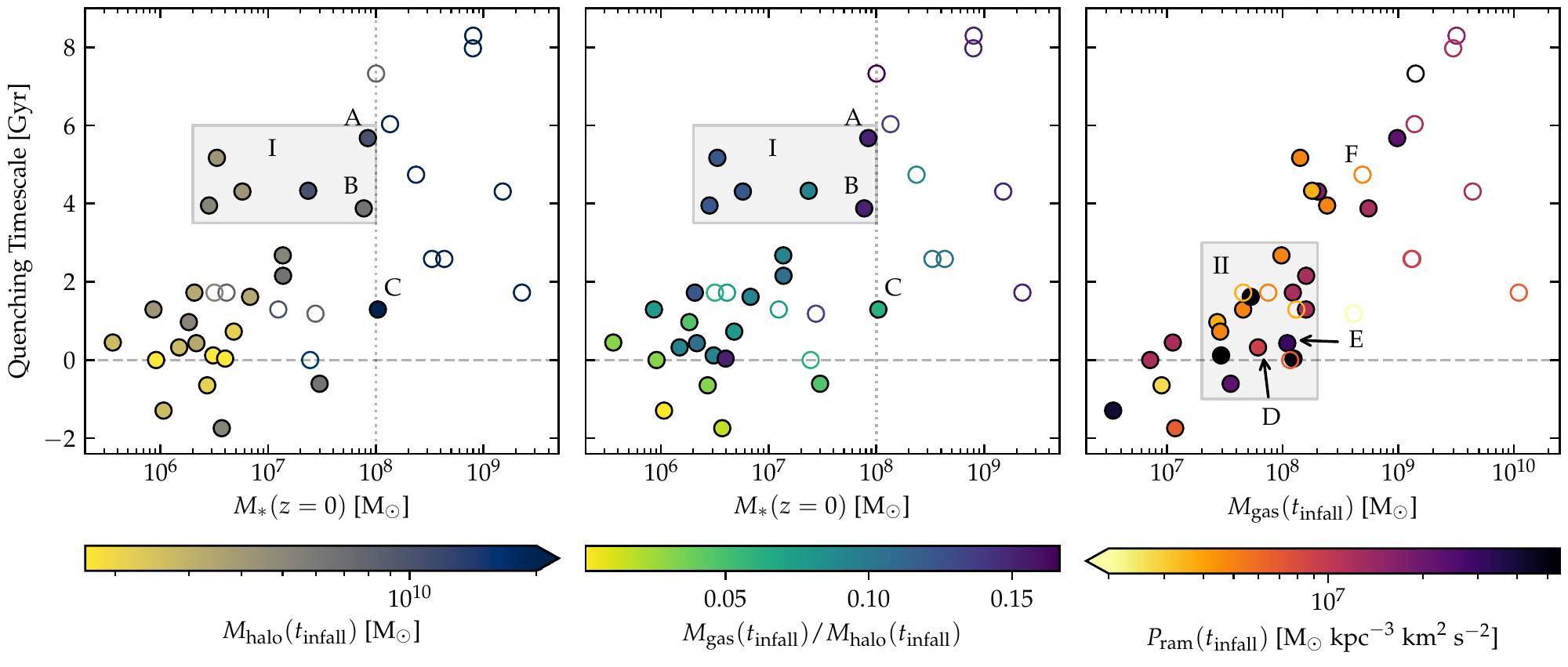}
\caption{
	Quenching timescale as it relates to stellar mass, gas mass, halo mass, and ram pressure. 
	In all three panels, the satellites shown are those that quench environmentally, defined as having $t_{\mathrm{quench}} < 9.5$ Gyr ago and a quenching timescale $> -2$ Gyr.
	Unlike in Figure~\ref{fig:timescales}, we show star-forming satellites, whose quenching timescales are lower limits, as open circles.
	Point labels are referenced in the main text. 
	\textbf{Left:} Quenching timescale as a function of $z=0$ stellar mass, with points colored by their halo mass at $t_{\mathrm{infall}}$.
	The dotted vertical line separates the stellar mass regimes in which we expect quenching to occur rapidly ($<10^8~\mathrm{M}_\sun$) versus inefficiently or not at all ($>10^8~\mathrm{M}_\sun$). 
	\textbf{Center:} Quenching timescale as a function of $z=0$ stellar mass, with points colored by their halo gas fraction ($M_{\mathrm{gas}}/M_{\mathrm{halo}}$) at infall. The dotted vertical line is the same as in the left panel.
	\textbf{Right:} Quenching timescale, here shown as a function of gas mass at $t_{\mathrm{infall}}$. The points are colored by the ram pressure felt by the satellite at infall.
	At a given stellar mass, longer quenching times are correlated with both larger halo masses and larger gas fractions.
	When, as in the right panel, the quenching times are plotted against the combination of these two parameters ($M_{\mathrm{gas}} = M_{\mathrm{halo}} \times \frac{M_{\mathrm{gas}}}{M_{\mathrm{halo}}}$), a strong positive correlation appears. 
	A secondary dependency of quenching times on ram pressure is also evident, with lower pressures resulting in longer quenching times.
}\label{fig:gasmass_vrel}
\end{figure*}

The quenching timescales of the Justice League satellites are generally consistent with those of LG satellites and can thus provide insight into the quenching processes at play. 
Timescales of $\lesssim 2$ Gyr for intermediate-mass satellites ($M_* \sim 10^{6-8}~\mathrm{M}_\sun$) imply the role of an efficient environmental quenching process at this mass range.
A likely candidate for this process is ram-pressure stripping, which has been shown to rapidly remove gas from infalling satellites \citep{slater_mass_2014, wheeler_surprising_2014, wetzel_rapid_2015} and is particularly efficient in combination with SNe feedback \citep{caproni_threedimensional_2015, bahe_star_2015, kazantzidis_effects_2017}.
However, our results do not show \textit{universally} rapid timescales: several satellites in this same mass range take $\sim 5$ Gyr to quench. 
To help explain this scatter in the context of quenching processes, we introduce Figure~\ref{fig:gasmass_vrel}, which shows the quenching timescale as it relates to halo mass, gas mass, and ram pressure at infall. 
In this plot, we show only those satellites that quench environmentally, defined as having $t_{\mathrm{quench}} < 9.5$ Gyr ago and a quenching timescale $> -2$ Gyr. 
In each panel, open circles show star-forming satellites whose quenching timescales are lower limits.

The left panel shows the quenching timescale vs.~stellar mass, with points colored by the satellite's halo mass at infall.
The center panel also shows the quenching timescale vs.~stellar mass, this time with points colored by the ``gas fraction,'' defined as the ratio of gas mass to halo mass, at infall, where gas mass is computed including all gas within $1~R_{\rm vir}$ (excluding subhalos).
These two panels differentiate between two factors that likely affect a satellite's quenching timescale: its ability to hold onto its gas (related to the halo mass) and the relative size of its gas reservoir (related to the gas fraction).

We see that at a given $z=0$ stellar mass, those satellites that take $\sim 5$ Gyr to quench (points in the shaded region labeled I) generally have larger halo masses and gas fractions than those that quench rapidly. 
However, these two factors are not always associated with each other. 
For example, we note an outlier (labeled C) which quenches rapidly despite having $M_* \sim 10^8~\mathrm{M}_\sun$. 
While the satellite has a large halo mass, it has a particularly low gas fraction, which distinguishes it from similarly massive satellites and helps explain its much shorter quenching timescale. 
Similarly, while the galaxies labeled A and B have similar stellar masses and gas fractions, the lower halo mass of galaxy B helps explain its $\sim 2$ Gyr shorter timescale. 

The right panel of Figure~\ref{fig:gasmass_vrel} combines the depth of the potential well ($M_{\mathrm{halo}}$) and the relative size of the gas reservoir ($M_{\mathrm{gas}}/M_{\mathrm{halo}}$) into a single parameter: the absolute gas mass at infall ($M_{\mathrm{gas}} = M_{\mathrm{halo}} \times \frac{M_{\mathrm{gas}}}{M_{\mathrm{halo}}}$).
The quenching timescale is shown versus $M_{\mathrm{gas}}$, and we see a strong correlation, as the satellites with quenching timescales of $\gtrsim 5$ Gyr generally have large gas masses. 
On this panel, points are colored by the ram pressure felt by the satellite at infall, $P_{\mathrm{ram}} = \rho_{\mathrm{CGM}}\, v_{\mathrm{rel}}^2$, where $\rho_{\mathrm{CGM}}$ is the density of the CGM and $v_{\mathrm{rel}}$ is the satellite's velocity relative to its host. 
We compute $\rho_{\rm CGM}$ from a spherically-averaged gas density profile, centered on the main halo, measured at the radial distance of the satellite. 
Under this definition, $\rho_{\rm CGM}(1~R_{\rm vir})$ decreases from $\sim 10^3~\mathrm{M}_\sun\,\mathrm{kpc}^{-3}$ at $z=1.5$ to $
\sim 10^2~\mathrm{M}_\sun\,\mathrm{kpc}^{-3}$ at $z=0$.

At intermediate gas masses, (points in the shaded region labeled II), the satellites with longer quenching timescales experience less ram pressure than others at the same mass. 
While additional quenching processes are likely at play, this dependence on $P_{\mathrm{ram}}$ suggests that ram-pressure stripping drives gas removal in these rapidly quenching satellites, reaffirming previous theoretical work \citep[e.g.][]{mayer_simultaneous_2006,simpson_quenching_2018, Tremmel2019} as well as interpretations of observations \citep{slater_mass_2014,fillingham_taking_2015}. 
Galaxies with even smaller gas reservoirs quench rapidly, regardless of the ram pressure they feel, implying that these galaxies have shallow gravitational potential wells, in addition to small amounts of gas.

\begin{figure*}
    \centering
    \includegraphics[width=0.9\linewidth]{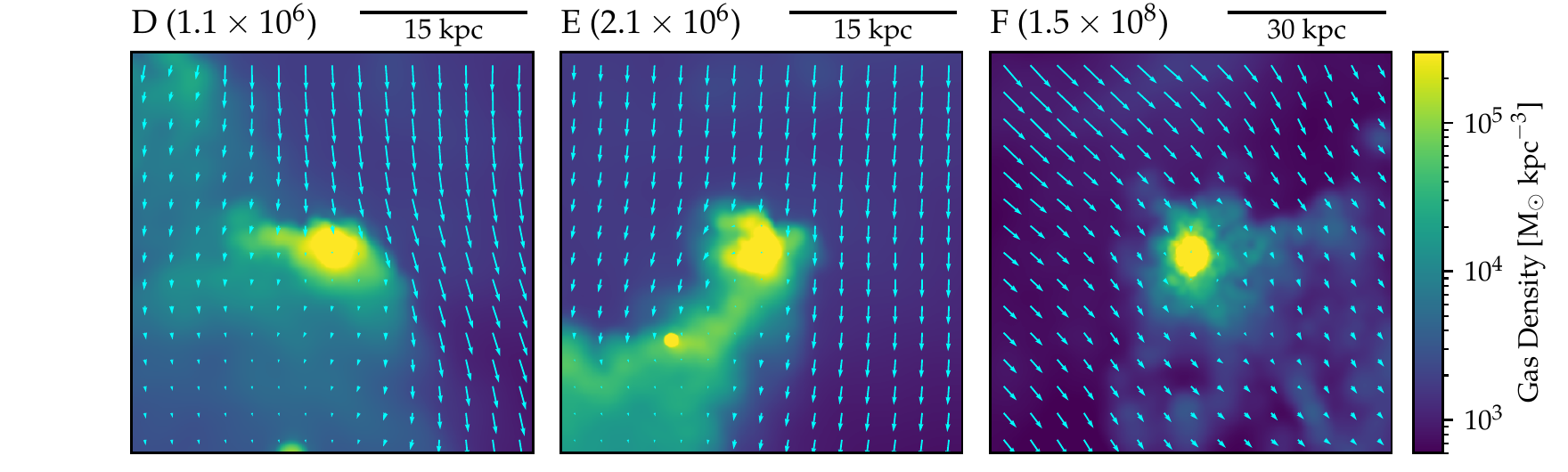}
    \caption{Line-of-sight averaged gas density maps for three satellites at infall. In each panel, the view is rotated such that the satellite's velocity relative to its host points in the positive $y$-direction. The velocity vectors thus show the velocity of the CGM relative to the satellite. 
    Each panel is annotated with a letter (see Figure~\ref{fig:gasmass_vrel}) and, in parentheses, the satellite's stellar mass at infall, in solar masses.
    For galaxy F, the velocity relative to the CGM is not in the same direction as the velocity relative to the central halo because the satellite infell as part of a group. 
    We see significant ram pressure trails in the low-mass satellites, highlighting the efficiency of ram pressure in this mass regime.}
    \label{fig:ram_pressure_images}
\end{figure*}

To further reaffirm the role of ram-pressure stripping in quenching intermediate-mass satellites, we introduce Figure~\ref{fig:ram_pressure_images}, which shows gas density maps in three satellites at their respective infall times. 
These satellites are labeled as points D, E, and F in Figure~\ref{fig:gasmass_vrel}, and vectors in Figure~\ref{fig:ram_pressure_images} indicate the velocity of the CGM relative to the satellite.
Galaxies D and E both quench rapidly and have gas masses in the regime where we see a strong trend with $P_{\mathrm{ram}}$, and indeed, they show strong trails of gas actively being stripped by ram pressure.
In contrast, galaxy F is a high-mass galaxy that is unquenched at $z=0$ despite having infell $\sim 5$ Gyr ago. 
It does not show evidence of significant ram-pressure stripping, as we see only a small trail of low-density gas opposite its direction of motion.

We note that $P_{\mathrm{ram}}(t_{\mathrm{infall}})$ does not account for the complexity of a satellite's orbit, which can be important when considering the effects of ram pressure. 
For example, we note an outlier in the right panel of Figure~\ref{fig:gasmass_vrel} with $M_{\mathrm{gas}}\sim 2\times10^8~\mathrm{M}_\sun$, a quenching timescale $>4$ Gyr, and $P_{\mathrm{ram}} \gtrsim 2\times10^7~\mathrm{M}_\sun~\mathrm{kpc}^{-3}~\mathrm{km}^2~\mathrm{s}^{-2}$. 
This satellite shows a unique SFH, oscillating between bursts of star-formation and periods of quiescence in the few Gyr after infall. 
Though its relative velocity is large, its radial velocity in the direction towards the host halo is small, suggesting that it is on a low-eccentricity orbit and does not move quickly move towards regions of increasing density. 
Similarly, the slightly longer-quenching galaxy in region II with $P_{\rm ram}(t_{\rm infall}) \sim 5\times10^6~\mathrm{M}_\sun~\mathrm{kpc}^{-3}~\mathrm{km}^2~\mathrm{s}^{-2}$ is a galaxy that is on a low-eccentricity orbit and does not experience a significant increase in ram pressure after infall.

Furthermore, considering orbital properties only at $t_{\mathrm{infall}}$ likely does not show the full picture for the satellites with longer quenching timescales. 
For rapidly quenching satellites, $P_{\mathrm{ram}}(t_{\mathrm{infall}})$ is an appropriate metric for assessing the role of ram pressure as $P_{\mathrm{ram}}(t_{\mathrm{infall}}) \approx P_{\mathrm{ram}}(t_{\mathrm{quench}})$, i.e.~the ram pressure does not change significantly over the quenching timescale.
However, for satellites that quench more slowly, the satellite's velocity and the environment through which it moves can evolve significantly. 
In particular, as the satellite approaches the central galaxy, its radial velocity and the local density will both increase, resulting in higher values of ram pressure.
Indeed, future work will explore the evolution of these satellites over their quenching timescales, and account for the local variations in the CGM density and temperature that can impact ram-pressure stripping.

Nevertheless, when examining the ram pressure at a single point in time, we argue that looking at the time of infall provides a reasonably fair comparison across satellites.
Satellites that experience higher ram pressure force at the time of infall, are generally on faster, more radial orbits or accreting onto more massive hosts and so will continue to experience higher ram pressure forces.
Additionally, since the local host CGM density changes slowly with radius, satellites accreting later and thus into a lower-density CGM will remain in lower-density material for a significant period of time.

We further argue that the fact that most satellites that quench environmentally do so near the time of infall indicates that examining the ram pressure at that time is appropriate for understanding quenching.
In contrast, examining the ram pressure at the time of {\em quenching} would, in addition to being undefined for the unquenched satellites, introduce its own set of biases, as longer-quenching satellites would have their ram pressure measured when they were on closer orbits to the main halo.

Finally, the long quenching timescales for satellites with $M_* \gtrsim  10^8~\mathrm{M}_\sun$ are consistent with those determined observationally \citep[e.g.][]{fillingham_taking_2015,phillips_mass_2015}, and imply that environmental quenching of these satellites is much less efficient.
Suppressed accretion of pristine gas, rather than ram-pressure stripping, may the dominant quenching mechanism in this mass regime \citep{wheeler_surprising_2014}. 
Evidence for the decreasing efficiency of ram-pressure stripping at these masses can be seen in the right panel of Figure~\ref{fig:gasmass_vrel}: most satellites that remain star-forming do so in spite of high ram pressure. 
It is clear that higher-mass satellites are resistant to ram-pressure stripping. 
Likewise, the lack of quenched galaxies with  $M_* \gtrsim  10^8~\mathrm{M}_\sun$ implies all other quenching processes in this mass range must also be slow to the extent they are active at all.
Indeed, our stellar mass threshold corresponds to $M_{\mathrm{halo}}(t_{\mathrm{infall}})\sim 10^{10}~\mathrm{M}_\odot$, consistent with previous theoretical estimates of the maximum halo mass at which ram pressure, aided by tidal stripping, is efficient \citep{mayer_simultaneous_2006}.

\section{Conclusions}\label{sec:conclusion}

Using the high-resolution DC Justice League simulations, we have explored the quenched fractions and quenching timescales of dwarf galaxies near Milky Way-mass hosts. 
Our conclusions are summarized as follows: 
\begin{enumerate}
	\item The Justice League simulations accurately reproduce observations of the dwarf galaxies in the LG, both with regards to satellite luminosity functions and quenched fractions across 5 dex in stellar mass. 
	While we see variability between individual simulations, this is largely consistent with observations of the LG and other nearby host-satellite systems.
	
	\item We find that intermediate-mass satellites ($M_* \sim 10^{6-8}~\mathrm{M}_\sun$) generally have short quenching timescales of $\lesssim 2$ Gyr, indicating that quenching is very efficient at this mass.
	However, several satellites in this mass range do not quench rapidly, instead continuing to form stars for as much as $5$ Gyr after infall. 
	These special cases generally have larger gas fractions at infall than similarly luminous satellites that quench rapidly. 
	
	\item We find a strong positive correlation between the quenching timescale and the satellite gas mass at infall. 
	For satellites that quench rapidly, the quenching timescale is also inversely correlated with the ram pressure felt by the satellite at infall. 
	These correlations suggests that ram-pressure stripping is the dominant quenching mechanism for dwarf satellites with $M_* \sim 10^{6-8}~\mathrm{M}_\sun$. The presence of ram pressure trails in satellites within this mass range further corroborates this finding.
	
	\item We find a $z=0$ stellar mass threshold of $\sim 10^8~\mathrm{M}_\sun$, above which infalling satellites are largely resistant to rapid environmental quenching and can remain star-forming to $z=0$ for as much as $8$ Gyr after infall.
	Only one of our satellites with $M_* > 10^8~\mathrm{M}_\sun$ is quenched and it has a mass of $1.04\times10^8~\mathrm{M}_\sun$. 
	Below this threshold, only five satellites remain star-forming, all of which fell into their host halo $\sim 1$--$2$ Gyr ago.  
	This may point to a characteristic mass scale at which the quenching process---and with it the quenching efficiency---shifts.
\end{enumerate}

There is much further work to be done in analyzing the Justice League suite of simulations.
While this work primarily focuses on the overall populations of dwarf satellites in the Justice League, future work will investigate the evolution of individual systems and their contribution to the CGM. 
At present, we are using particle-tracking to investigate the spatial distribution, temperature, and density of satellite gas post-infall.
Preliminary results indicate that gas stripping timescales are highly correlated with the quenching timescales of intermediate-mass satellites, though dependent on gas phase, and these results will appear in a future paper. 
While it remains unclear the precise role that each quenching mechanism may play in the quenching of dwarf satellites, the Justice League simulations provide a promising route by which to better understand these processes.

Missing in this analysis are the quenched fractions and quenching timescales of ultra-faint dwarfs (UFDs; $M_* < 10^5~\mathrm{M}_\Sun$). 
Although the Justice League simulations used in this work are run at ``near-mint'' resolution, these dwarfs approach our resolution limit and may be subject to additional uncertainty in the modeling of baryonic physics. 
Additional Justice League runs at even higher ``mint'' resolution have been completed, allowing us to more throughly explore UFD quenching and the effects of reionization on these faint galaxies \citep{applebaum_ultrafaint_2020}.

\acknowledgments

The authors are grateful to the anonymous referee for thoughtful critique that improved the quality of this paper.
The authors additionally thank Arif Babul, Jillian Bellovary, Pedro Capelo, Lucio Mayer, and Tom Quinn for thorough comments and suggestions in the late stages of this project.  
H.B.A. acknowledges support from Grinnell College through the Mentored Advanced Project (MAP) program.
This work is supported by the U.S.~NSF under CAREER grant AST-1848107. 
A.M.B. is partially supported by NSF grant AST-1813871. 
Resources supporting this work were provided by the NASA High-End Computing (HEC) Program through the NASA Advanced Supercomputing (NAS) Division at Ames Research Center. 

\software{\textsc{Python}, \textsc{numpy} \citep{walt_numpy_2011}, \textsc{matplotlib} \citep{hunter_matplotlib_2007}, \textsc{Jupyter}, \ChaNGa \citep{menon_adaptive_2015}, \textsc{Sunrise} \citep{jonsson_highresolution_2010}, \textsc{pynbody} \citep{pontzen_pynbody_2013}, \textsc{tangos} \citep{pontzen_tangos_2018}, PARSEC \citep{Bressan2012}.}

\bibliographystyle{aasjournal} 
\bibliography{timescales_paper}

\begin{thebibliography}{}
\expandafter\ifx\csname natexlab\endcsname\relax\def\natexlab#1{#1}\fi
\providecommand{\url}[1]{\href{#1}{#1}}
\providecommand{\dodoi}[1]{doi:~\href{http://doi.org/#1}{\nolinkurl{#1}}}
\providecommand{\doeprint}[1]{\href{http://ascl.net/#1}{\nolinkurl{http://ascl.net/#1}}}
\providecommand{\doarXiv}[1]{\href{https://arxiv.org/abs/#1}{\nolinkurl{https://arxiv.org/abs/#1}}}

\bibitem[{{Applebaum} {et~al.}(2021){Applebaum}, {Brooks}, {Christensen},
  {Munshi}, {Quinn}, {Shen}, \& {Tremmel}}]{applebaum_ultrafaint_2020}
{Applebaum}, E., {Brooks}, A.~M., {Christensen}, C.~R., {et~al.} 2021, \apj,
  906, 96, \dodoi{10.3847/1538-4357/abcafa}

\bibitem[{Bah{\'e} \& McCarthy(2015)}]{bahe_star_2015}
Bah{\'e}, Y.~M., \& McCarthy, I.~G. 2015, MNRAS, 447, 969,
  \dodoi{10.1093/mnras/stu2293}

\bibitem[{Bah{\'e} {et~al.}(2013)Bah{\'e}, McCarthy, Balogh, \&
  Font}]{bahe_why_2013}
Bah{\'e}, Y.~M., McCarthy, I.~G., Balogh, M.~L., \& Font, A.~S. 2013, MNRAS,
  430, 3017, \dodoi{10.1093/mnras/stu2293}

\bibitem[{Bellovary {et~al.}(2011)Bellovary, Volonteri, Governato, Shen, Quinn,
  \& Wadsley}]{bellovary_first_2011}
Bellovary, J., Volonteri, M., Governato, F., {et~al.} 2011, ApJ, 742, 13,
  \dodoi{10.1088/0004-637X/742/1/13}

\bibitem[{Bellovary {et~al.}(2019)Bellovary, Cleary, Munshi, Tremmel,
  Christensen, Brooks, \& Quinn}]{bellovary_multimessenger_2019}
Bellovary, J.~M., Cleary, C.~E., Munshi, F., {et~al.} 2019, \mnras, 482, 2913,
  \dodoi{10.1093/mnras/sty2842}

\bibitem[{{Ben{\'i}tez-Llambay} {et~al.}(2015){Ben{\'i}tez-Llambay}, Navarro,
  Abadi, Gottl{\"o}ber, Yepes, Hoffman, \&
  Steinmetz}]{benitez-llambay_imprint_2015}
{Ben{\'i}tez-Llambay}, A., Navarro, J.~F., Abadi, M.~G., {et~al.} 2015, \mnras,
  450, 4207, \dodoi{10.1093/mnras/stv925}

\bibitem[{{Bennet} {et~al.}(2019){Bennet}, {Sand}, {Crnojevi{\'c}}, {Spekkens},
  {Karunakaran}, {Zaritsky}, \& {Mutlu-Pakdil}}]{bennet_m101_2019}
{Bennet}, P., {Sand}, D.~J., {Crnojevi{\'c}}, D., {et~al.} 2019, \apj, 885,
  153, \dodoi{10.3847/1538-4357/ab46ab}

\bibitem[{{Bressan} {et~al.}(2012){Bressan}, {Marigo}, {Girardi}, {Salasnich},
  {Dal Cero}, {Rubele}, \& {Nanni}}]{Bressan2012}
{Bressan}, A., {Marigo}, P., {Girardi}, L., {et~al.} 2012, \mnras, 427, 127,
  \dodoi{10.1111/j.1365-2966.2012.21948.x}

\bibitem[{Brooks {et~al.}(2013)Brooks, Kuhlen, Zolotov, \&
  Hooper}]{brooks_baryonic_2013}
Brooks, A.~M., Kuhlen, M., Zolotov, A., \& Hooper, D. 2013, ApJ, 765, 22,
  \dodoi{10.1088/0004-637X/765/1/22}

\bibitem[{Brooks \& Zolotov(2014)}]{brooks_why_2014}
Brooks, A.~M., \& Zolotov, A. 2014, ApJ, 786, 87,
  \dodoi{10.1088/0004-637X/786/2/87}

\bibitem[{Brown {et~al.}(2014)Brown, Tumlinson, Geha, Simon, Vargas,
  VandenBerg, Kirby, Kalirai, Avila, Gennaro, Ferguson, Mu{\~n}oz,
  Guhathakurta, \& Renzini}]{brown_quenching_2014}
Brown, T.~M., Tumlinson, J., Geha, M., {et~al.} 2014, ApJ, 796, 91,
  \dodoi{10.1088/0004-637X/796/2/91}

\bibitem[{{Buck} {et~al.}(2019){Buck}, {Macci{\`o}}, {Dutton}, {Obreja}, \&
  {Frings}}]{Buck_2019}
{Buck}, T., {Macci{\`o}}, A.~V., {Dutton}, A.~A., {Obreja}, A., \& {Frings}, J.
  2019, \mnras, 483, 1314, \dodoi{10.1093/mnras/sty2913}

\bibitem[{Caproni {et~al.}(2015)Caproni, Lanfranchi, {da Silva}, \&
  {Falceta-Gon{\c c}alves}}]{caproni_threedimensional_2015}
Caproni, A., Lanfranchi, G.~A., {da Silva}, A.~L., \& {Falceta-Gon{\c c}alves},
  D. 2015, ApJ, 805, 109, \dodoi{10.1088/0004-637X/805/2/109}

\bibitem[{{Carlsten} {et~al.}(2020){Carlsten}, {Greene}, {Peter}, {Greco}, \&
  {Beaton}}]{carlsten2020luminosity}
{Carlsten}, S.~G., {Greene}, J.~E., {Peter}, A. H.~G., {Greco}, J.~P., \&
  {Beaton}, R.~L. 2020, \apj, 902, 124, \dodoi{10.3847/1538-4357/abb60b}

\bibitem[{Chiboucas {et~al.}(2013)Chiboucas, Jacobs, Tully, \&
  Karachentsev}]{chiboucas_confirmation_2013}
Chiboucas, K., Jacobs, B.~A., Tully, R.~B., \& Karachentsev, I.~D. 2013, ApJ,
  146, 126, \dodoi{10.1088/0004-6256/146/5/126}

\bibitem[{Christensen {et~al.}(2012)Christensen, Quinn, Governato, Stilp, Shen,
  \& Wadsley}]{christensen_implementing_2012}
Christensen, C., Quinn, T., Governato, F., {et~al.} 2012, MNRAS, 425, 3058,
  \dodoi{10.1111/j.1365-2966.2012.21628.x}

\bibitem[{Christensen {et~al.}(2016)Christensen, Dav{\'e}, Governato, Pontzen,
  Brooks, Munshi, Quinn, \& Wadsley}]{christensen_innout_2016}
Christensen, C.~R., Dav{\'e}, R., Governato, F., {et~al.} 2016, ApJ, 824, 57,
  \dodoi{10.3847/0004-637X/824/1/57}

\bibitem[{Crnojevi{\'c} {et~al.}(2019)Crnojevi{\'c}, Sand, Bennet, Pasetto,
  Spekkens, Caldwell, Guhathakurta, McLeod, Seth, Simon, Strader, \&
  Toloba}]{crnojevic_faint_2019}
Crnojevi{\'c}, D., Sand, D.~J., Bennet, P., {et~al.} 2019, ApJ, 872, 80,
  \dodoi{10.3847/1538-4357/aafbe7}

\bibitem[{{de Vaucouleurs} {et~al.}(1991){de Vaucouleurs}, {de Vaucouleurs},
  Corwin, Buta, Paturel, \& Fouqu{\'e}}]{devaucouleurs_third_1991}
{de Vaucouleurs}, G., {de Vaucouleurs}, A., Corwin, Jr., H.~G., {et~al.} 1991,
  Third {{Reference Catalogue}} of {{Bright Galaxies}}. ({New York, NY (USA)}:
  {Springer})

\bibitem[{Digby {et~al.}(2019)Digby, Navarro, Fattahi, Simpson, Oman, Gomez,
  Frenk, Grand, \& Pakmor}]{digby_star_2019}
Digby, R., Navarro, J.~F., Fattahi, A., {et~al.} 2019, MNRAS, 485, 5423,
  \dodoi{10.1093/mnras/stz745}

\bibitem[{Emerick {et~al.}(2016)Emerick, Mac~Low, Grcevich, \&
  Gatto}]{emerick_gas_2016}
Emerick, A., Mac~Low, M.-M., Grcevich, J., \& Gatto, A. 2016, ApJ, 826, 148,
  \dodoi{10.3847/0004-637X/826/2/148}

\bibitem[{Fillingham {et~al.}(2015)Fillingham, Cooper, Wheeler,
  {Garrison-Kimmel}, {Boylan-Kolchin}, \& Bullock}]{fillingham_taking_2015}
Fillingham, S.~P., Cooper, M.~C., Wheeler, C., {et~al.} 2015, MNRAS, 454, 2039,
  \dodoi{10.1093/mnras/stv2058}

\bibitem[{Fillingham {et~al.}(2019)Fillingham, Cooper, Kelley,
  Rodriguez~Wimberly, {Boylan-Kolchin}, Bullock, {Garrison-Kimmel}, Pawlowski,
  \& Wheeler}]{fillingham_characterizing_2019}
Fillingham, S.~P., Cooper, M.~C., Kelley, T., {et~al.} 2019, arXiv e-prints,
  arXiv:1906.04180

\bibitem[{{Garrison-Kimmel} {et~al.}(2014){Garrison-Kimmel}, {Boylan-Kolchin},
  {Bullock}, \& {Lee}}]{Garrison-Kimmel2014}
{Garrison-Kimmel}, S., {Boylan-Kolchin}, M., {Bullock}, J.~S., \& {Lee}, K.
  2014, \mnras, 438, 2578, \dodoi{10.1093/mnras/stt2377}

\bibitem[{{Garrison-Kimmel} {et~al.}(2017){Garrison-Kimmel}, {Wetzel},
  {Bullock}, {Hopkins}, {Boylan-Kolchin}, {Faucher-Gigu{\`e}re}, {Kere{\v{s}}},
  {Quataert}, {Sanderson}, {Graus}, \& {Kelley}}]{garrison-kimmel_2017}
{Garrison-Kimmel}, S., {Wetzel}, A., {Bullock}, J.~S., {et~al.} 2017, \mnras,
  471, 1709, \dodoi{10.1093/mnras/stx1710}

\bibitem[{{Garrison-Kimmel} {et~al.}(2019){Garrison-Kimmel}, Wetzel, Hopkins,
  Sanderson, {El-Badry}, Graus, Chan, Feldmann, {Boylan-Kolchin}, Hayward,
  Bullock, Fitts, Samuel, Wheeler, Keres, \&
  {Faucher-Giguere}}]{garrison-kimmel_star_2019}
{Garrison-Kimmel}, S., Wetzel, A., Hopkins, P.~F., {et~al.} 2019, arXiv
  e-prints, arXiv:1903.10515

\bibitem[{Geha {et~al.}(2012)Geha, Blanton, Yan, \& Tinker}]{geha_stellar_2012}
Geha, M., Blanton, M.~R., Yan, R., \& Tinker, J.~L. 2012, ApJ, 757, 85,
  \dodoi{10.1088/0004-637X/757/1/85}

\bibitem[{Geha {et~al.}(2017)Geha, Wechsler, Mao, Tollerud, Weiner, Bernstein,
  Hoyle, Marchi, Marshall, Mu{\~n}oz, \& Lu}]{geha_saga_2017}
Geha, M., Wechsler, R.~H., Mao, Y.-Y., {et~al.} 2017, ApJ, 847, 4,
  \dodoi{10.3847/1538-4357/aa8626}

\bibitem[{Governato {et~al.}(2010)Governato, Brook, Mayer, Brooks, Rhee,
  Wadsley, Jonsson, Willman, Stinson, Quinn, \&
  Madau}]{governato_bulgeless_2010}
Governato, F., Brook, C., Mayer, L., {et~al.} 2010, Nature, 463, 203,
  \dodoi{10.1038/nature08640}

\bibitem[{Grcevich \& Putman(2009)}]{grcevich_local_2009}
Grcevich, J., \& Putman, M.~E. 2009, ApJ, 696, 385,
  \dodoi{10.1088/0004-637X/696/1/385}

\bibitem[{Haardt \& Madau(2012)}]{haardt_radiative_2012}
Haardt, F., \& Madau, P. 2012, ApJ, 746, 125,
  \dodoi{10.1088/0004-637X/746/2/125}

\bibitem[{Hopkins {et~al.}(2014)Hopkins, Kere{\v s}, O{\~n}orbe,
  {Faucher-Gigu{\`e}re}, Quataert, Murray, \& Bullock}]{hopkins_galaxies_2014}
Hopkins, P.~F., Kere{\v s}, D., O{\~n}orbe, J., {et~al.} 2014, \mnras, 445,
  581, \dodoi{10.1093/mnras/stu1738}

\bibitem[{Huang {et~al.}(2012)Huang, Haynes, Giovanelli, Brinchmann, Stierwalt,
  \& Neff}]{huang_gas_2012}
Huang, S., Haynes, M.~P., Giovanelli, R., {et~al.} 2012, The Astronomical
  Journal, 143, 133, \dodoi{10.1088/0004-6256/143/6/133}

\bibitem[{Hunter(2007)}]{hunter_matplotlib_2007}
Hunter, J.~D. 2007, Computing in Science Engineering, 9, 90,
  \dodoi{10.1109/MCSE.2007.55}

\bibitem[{Jonsson {et~al.}(2010)Jonsson, Groves, \&
  Cox}]{jonsson_highresolution_2010}
Jonsson, P., Groves, B.~A., \& Cox, T.~J. 2010, \mnras, 403, 17,
  \dodoi{10.1111/j.1365-2966.2009.16087.x}

\bibitem[{Jung {et~al.}(2018)Jung, Choi, Wong, Kimm, Chung, \&
  Yi}]{jung_origin_2018}
Jung, S.~L., Choi, H., Wong, O.~I., {et~al.} 2018, ApJ, 865, 156,
  \dodoi{10.3847/1538-4357/aadda2}

\bibitem[{Kallivayalil {et~al.}(2013)Kallivayalil, {van der Marel}, Besla,
  Anderson, \& Alcock}]{kallivayalil_thirdepoch_2013}
Kallivayalil, N., {van der Marel}, R.~P., Besla, G., Anderson, J., \& Alcock,
  C. 2013, ApJ, 764, 161, \dodoi{10.1088/0004-637X/764/2/161}

\bibitem[{{Karachentsev} \& {Kudrya}(2014)}]{Karachentsev2014}
{Karachentsev}, I.~D., \& {Kudrya}, Y.~N. 2014, \aj, 148, 50,
  \dodoi{10.1088/0004-6256/148/3/50}

\bibitem[{Karachentsev {et~al.}(2013)Karachentsev, Makarov, \&
  Kaisina}]{karachentsev_updated_2013}
Karachentsev, I.~D., Makarov, D.~I., \& Kaisina, E.~I. 2013, AJ, 145, 101,
  \dodoi{10.1088/0004-6256/145/4/101}

\bibitem[{Katz {et~al.}(2020)Katz, Ramsoy, Rosdahl, Kimm, Blaizot, Haehnelt,
  {Michel-Dansac}, Garel, Laigle, Devriendt, \& Slyz}]{katz_how_2020}
Katz, H., Ramsoy, M., Rosdahl, J., {et~al.} 2020, \mnras, 494, 2200,
  \dodoi{10.1093/mnras/staa639}

\bibitem[{Katz \& White(1993)}]{katz_hierarchical_1993}
Katz, N., \& White, S. D.~M. 1993, ApJ, 412, 455, \dodoi{10.1086/172935}

\bibitem[{Kazantzidis {et~al.}(2017)Kazantzidis, Mayer, Callegari, Dotti, \&
  Moustakas}]{kazantzidis_effects_2017}
Kazantzidis, S., Mayer, L., Callegari, S., Dotti, M., \& Moustakas, L.~A. 2017,
  ApJ, 836, L13, \dodoi{10.3847/2041-8213/aa5b8f}

\bibitem[{{Keller} {et~al.}(2014){Keller}, {Wadsley}, {Benincasa}, \&
  {Couchman}}]{Keller2014}
{Keller}, B.~W., {Wadsley}, J., {Benincasa}, S.~M., \& {Couchman}, H.~M.~P.
  2014, \mnras, 442, 3013, \dodoi{10.1093/mnras/stu1058}

\bibitem[{Klypin {et~al.}(1999)Klypin, Kravtsov, Valenzuela, \&
  Prada}]{klypin_where_1999}
Klypin, A., Kravtsov, A.~V., Valenzuela, O., \& Prada, F. 1999, ApJ, 522, 82,
  \dodoi{10.1086/307643}

\bibitem[{Knollmann \& Knebe(2009)}]{knollmann_ahf_2009}
Knollmann, S.~R., \& Knebe, A. 2009, ApJS, 182, 608,
  \dodoi{10.1088/0067-0049/182/2/608}

\bibitem[{Kroupa(2002)}]{kroupa_initial_2002}
Kroupa, P. 2002, Science, 295, 82, \dodoi{10.1126/science.1067524}

\bibitem[{{Mao} {et~al.}(2020){Mao}, {Geha}, {Wechsler}, {Weiner}, {Tollerud},
  {Nadler}, \& {Kallivayalil}}]{mao_saga_2020}
{Mao}, Y.-Y., {Geha}, M., {Wechsler}, R.~H., {et~al.} 2020, arXiv e-prints,
  arXiv:2008.12783.
\newblock \doarXiv{2008.12783}

\bibitem[{Mayer {et~al.}(2006)Mayer, Mastropietro, Wadsley, Stadel, \&
  Moore}]{mayer_simultaneous_2006}
Mayer, L., Mastropietro, C., Wadsley, J., Stadel, J., \& Moore, B. 2006, MNRAS,
  369, 1021, \dodoi{10.1111/j.1365-2966.2006.10403.x}

\bibitem[{McConnachie(2012)}]{mcconnachie_observed_2012}
McConnachie, A.~W. 2012, AJ, 144, 4, \dodoi{10.1088/0004-6256/144/1/4}

\bibitem[{McGee {et~al.}(2014)McGee, Bower, \&
  Balogh}]{mcgee_overconsumption_2014}
McGee, S.~L., Bower, R.~G., \& Balogh, M.~L. 2014, MNRAS, 442, L105,
  \dodoi{10.1093/mnrasl/slu066}

\bibitem[{McKee \& Ostriker(1977)}]{mckee_theory_1977}
McKee, C.~F., \& Ostriker, J.~P. 1977, ApJ, 218, 148, \dodoi{10.1086/155667}

\bibitem[{Menon {et~al.}(2015)Menon, Wesolowski, Zheng, Jetley, Kale, Quinn, \&
  Governato}]{menon_adaptive_2015}
Menon, H., Wesolowski, L., Zheng, G., {et~al.} 2015, CompAC, 2, 1,
  \dodoi{10.1186/s40668-015-0007-9}

\bibitem[{{Murakami} \& {Babul}(1999)}]{murakami_babul_1999}
{Murakami}, I., \& {Babul}, A. 1999, \mnras, 309, 161,
  \dodoi{10.1046/j.1365-8711.1999.02810.x}

\bibitem[{Nickerson {et~al.}(2013)Nickerson, Stinson, Couchman, Bailin, \&
  Wadsley}]{nickerson_luminosity_2013}
Nickerson, S., Stinson, G., Couchman, H. M.~P., Bailin, J., \& Wadsley, J.
  2013, MNRAS, 429, 452, \dodoi{10.1093/mnras/sts348}

\bibitem[{{O{\~n}orbe} {et~al.}(2017){O{\~n}orbe}, {Hennawi}, \&
  {Luki{\'c}}}]{Onorbe2017}
{O{\~n}orbe}, J., {Hennawi}, J.~F., \& {Luki{\'c}}, Z. 2017, \apj, 837, 106,
  \dodoi{10.3847/1538-4357/aa6031}

\bibitem[{Pallero {et~al.}(2019)Pallero, G{\'o}mez, Padilla, {Torres-Flores},
  Demarco, Cerulo, \& {Olave-Rojas}}]{pallero_tracing_2019}
Pallero, D., G{\'o}mez, F.~A., Padilla, N.~D., {et~al.} 2019, MNRAS, 488, 847,
  \dodoi{10.1093/mnras/stz1745}

\bibitem[{Phillips {et~al.}(2015)Phillips, Wheeler, Cooper, {Boylan-Kolchin},
  Bullock, \& Tollerud}]{phillips_mass_2015}
Phillips, J.~I., Wheeler, C., Cooper, M.~C., {et~al.} 2015, MNRAS, 447, 698,
  \dodoi{10.1093/mnras/stu2192}

\bibitem[{{Planck Collaboration} {et~al.}(2016){Planck Collaboration}, {Ade},
  {Aghanim}, {Arnaud}, {Ashdown}, {Aumont}, {Baccigalupi}, {Banday},
  {Barreiro}, {Bartlett}, {Bartolo}, {Battaner}, {Battye}, {Benabed},
  {Beno{\^\i}t}, {Benoit-L{\'e}vy}, {Bernard}, {Bersanelli}, {Bielewicz},
  {Bock}, {Bonaldi}, {Bonavera}, {Bond}, {Borrill}, {Bouchet}, {Boulanger},
  {Bucher}, {Burigana}, {Butler}, {Calabrese}, {Cardoso}, {Catalano},
  {Challinor}, {Chamballu}, {Chary}, {Chiang}, {Chluba}, {Christensen},
  {Church}, {Clements}, {Colombi}, {Colombo}, {Combet}, {Coulais}, {Crill},
  {Curto}, {Cuttaia}, {Danese}, {Davies}, {Davis}, {de Bernardis}, {de Rosa},
  {de Zotti}, {Delabrouille}, {D{\'e}sert}, {Di Valentino}, {Dickinson},
  {Diego}, {Dolag}, {Dole}, {Donzelli}, {Dor{\'e}}, {Douspis}, {Ducout},
  {Dunkley}, {Dupac}, {Efstathiou}, {Elsner}, {En{\ss}lin}, {Eriksen},
  {Farhang}, {Fergusson}, {Finelli}, {Forni}, {Frailis}, {Fraisse},
  {Franceschi}, {Frejsel}, {Galeotta}, {Galli}, {Ganga}, {Gauthier}, {Gerbino},
  {Ghosh}, {Giard}, {Giraud-H{\'e}raud}, {Giusarma}, {Gjerl{\o}w},
  {Gonz{\'a}lez-Nuevo}, {G{\'o}rski}, {Gratton}, {Gregorio}, {Gruppuso},
  {Gudmundsson}, {Hamann}, {Hansen}, {Hanson}, {Harrison}, {Helou},
  {Henrot-Versill{\'e}}, {Hern{\'a}ndez-Monteagudo}, {Herranz}, {Hildebrand t},
  {Hivon}, {Hobson}, {Holmes}, {Hornstrup}, {Hovest}, {Huang}, {Huffenberger},
  {Hurier}, {Jaffe}, {Jaffe}, {Jones}, {Juvela}, {Keih{\"a}nen}, {Keskitalo},
  {Kisner}, {Kneissl}, {Knoche}, {Knox}, {Kunz}, {Kurki-Suonio}, {Lagache},
  {L{\"a}hteenm{\"a}ki}, {Lamarre}, {Lasenby}, {Lattanzi}, {Lawrence}, {Leahy},
  {Leonardi}, {Lesgourgues}, {Levrier}, {Lewis}, {Liguori}, {Lilje},
  {Linden-V{\o}rnle}, {L{\'o}pez-Caniego}, {Lubin}, {Mac{\'\i}as-P{\'e}rez},
  {Maggio}, {Maino}, {Mandolesi}, {Mangilli}, {Marchini}, {Maris}, {Martin},
  {Martinelli}, {Mart{\'\i}nez-Gonz{\'a}lez}, {Masi}, {Matarrese}, {McGehee},
  {Meinhold}, {Melchiorri}, {Melin}, {Mendes}, {Mennella}, {Migliaccio},
  {Millea}, {Mitra}, {Miville-Desch{\^e}nes}, {Moneti}, {Montier}, {Morgante},
  {Mortlock}, {Moss}, {Munshi}, {Murphy}, {Naselsky}, {Nati}, {Natoli},
  {Netterfield}, {N{\o}rgaard-Nielsen}, {Noviello}, {Novikov}, {Novikov},
  {Oxborrow}, {Paci}, {Pagano}, {Pajot}, {Paladini}, {Paoletti}, {Partridge},
  {Pasian}, {Patanchon}, {Pearson}, {Perdereau}, {Perotto}, {Perrotta},
  {Pettorino}, {Piacentini}, {Piat}, {Pierpaoli}, {Pietrobon}, {Plaszczynski},
  {Pointecouteau}, {Polenta}, {Popa}, {Pratt}, {Pr{\'e}zeau}, {Prunet},
  {Puget}, {Rachen}, {Reach}, {Rebolo}, {Reinecke}, {Remazeilles}, {Renault},
  {Renzi}, {Ristorcelli}, {Rocha}, {Rosset}, {Rossetti}, {Roudier},
  {Rouill{\'e} d'Orfeuil}, {Rowan-Robinson}, {Rubi{\~n}o-Mart{\'\i}n},
  {Rusholme}, {Said}, {Salvatelli}, {Salvati}, {Sandri}, {Santos},
  {Savelainen}, {Savini}, {Scott}, {Seiffert}, {Serra}, {Shellard}, {Spencer},
  {Spinelli}, {Stolyarov}, {Stompor}, {Sudiwala}, {Sunyaev}, {Sutton},
  {Suur-Uski}, {Sygnet}, {Tauber}, {Terenzi}, {Toffolatti}, {Tomasi},
  {Tristram}, {Trombetti}, {Tucci}, {Tuovinen}, {T{\"u}rler}, {Umana},
  {Valenziano}, {Valiviita}, {Van Tent}, {Vielva}, {Villa}, {Wade}, {Wandelt},
  {Wehus}, {White}, {White}, {Wilkinson}, {Yvon}, {Zacchei}, \&
  {Zonca}}]{planck_2016}
{Planck Collaboration}, {Ade}, P.~A.~R., {Aghanim}, N., {et~al.} 2016, \aap,
  594, A13, \dodoi{10.1051/0004-6361/201525830}

\bibitem[{Pontzen {et~al.}(2013)Pontzen, Ro{\v s}kar, Stinson, \&
  Woods}]{pontzen_pynbody_2013}
Pontzen, A., Ro{\v s}kar, R., Stinson, G., \& Woods, R. 2013, Astrophysics
  Source Code Library, ascl:1305.002

\bibitem[{Pontzen \& Tremmel(2018)}]{pontzen_tangos_2018}
Pontzen, A., \& Tremmel, M. 2018, ApJS, 237, 23,
  \dodoi{10.3847/1538-4365/aac832}

\bibitem[{{Quilis} {et~al.}(2000){Quilis}, {Moore}, \& {Bower}}]{Quilis2000}
{Quilis}, V., {Moore}, B., \& {Bower}, R. 2000, Science, 288, 1617,
  \dodoi{10.1126/science.288.5471.1617}

\bibitem[{{Richings} {et~al.}(2020){Richings}, {Frenk}, {Jenkins}, {Robertson},
  {Fattahi}, {Grand}, {Navarro}, {Pakmor}, {Gomez}, {Marinacci}, \&
  {Oman}}]{Richings_2020}
{Richings}, J., {Frenk}, C., {Jenkins}, A., {et~al.} 2020, \mnras, 492, 5780,
  \dodoi{10.1093/mnras/stz3448}

\bibitem[{{Riley} {et~al.}(2019){Riley}, {Fattahi}, {Pace}, {Strigari},
  {Frenk}, {G{\'o}mez}, {Grand}, {Marinacci}, {Navarro}, {Pakmor}, {Simpson},
  \& {White}}]{Riley_2019}
{Riley}, A.~H., {Fattahi}, A., {Pace}, A.~B., {et~al.} 2019, \mnras, 486, 2679,
  \dodoi{10.1093/mnras/stz973}

\bibitem[{Ritchie \& Thomas(2001)}]{ritchie_multiphase_2001}
Ritchie, B.~W., \& Thomas, P.~A. 2001, MNRAS, 323, 743,
  \dodoi{10.1046/j.1365-8711.2001.04268.x}

\bibitem[{Rodriguez~Wimberly {et~al.}(2019)Rodriguez~Wimberly, Cooper,
  Fillingham, {Boylan-Kolchin}, Bullock, \&
  {Garrison-Kimmel}}]{rodriguezwimberly_suppression_2019}
Rodriguez~Wimberly, M.~K., Cooper, M.~C., Fillingham, S.~P., {et~al.} 2019,
  MNRAS, 483, 4031, \dodoi{10.1093/mnras/sty3357}

\bibitem[{Samuel {et~al.}(2020)Samuel, Wetzel, Tollerud, {Garrison-Kimmel},
  Loebman, {El-Badry}, Hopkins, {Boylan-Kolchin}, {Faucher-Gigu{\`e}re},
  Bullock, Benincasa, \& Bailin}]{samuel_profile_2020}
Samuel, J., Wetzel, A., Tollerud, E., {et~al.} 2020, \mnras, 491, 1471,
  \dodoi{10.1093/mnras/stz3054}

\bibitem[{Shen {et~al.}(2010)Shen, Wadsley, \& Stinson}]{shen_enrichment_2010}
Shen, S., Wadsley, J., \& Stinson, G. 2010, MNRAS, 407, 1581,
  \dodoi{10.1111/j.1365-2966.2010.17047.x}

\bibitem[{Simpson {et~al.}(2018)Simpson, Grand, G{\'o}mez, Marinacci, Pakmor,
  Springel, Campbell, \& Frenk}]{simpson_quenching_2018}
Simpson, C.~M., Grand, R. J.~J., G{\'o}mez, F.~A., {et~al.} 2018, MNRAS, 478,
  548, \dodoi{10.1093/mnras/sty774}

\bibitem[{Slater \& Bell(2013)}]{slater_confronting_2013}
Slater, C.~T., \& Bell, E.~F. 2013, ApJ, 773, 17,
  \dodoi{10.1088/0004-637X/773/1/17}

\bibitem[{Slater \& Bell(2014)}]{slater_mass_2014}
---. 2014, ApJ, 792, 141, \dodoi{10.1088/0004-637X/792/2/141}

\bibitem[{Smercina {et~al.}(2018)Smercina, Bell, Price, D'Souza, Slater,
  Bailin, Monachesi, \& Nidever}]{smercina_lonely_2018}
Smercina, A., Bell, E.~F., Price, P.~A., {et~al.} 2018, ApJ, 863, 152,
  \dodoi{10.3847/1538-4357/aad2d6}

\bibitem[{Smercina {et~al.}(2017)Smercina, Bell, Slater, Price, Bailin, \&
  Monachesi}]{smercina_d1005_2017}
Smercina, A., Bell, E.~F., Slater, C.~T., {et~al.} 2017, ApJ, 843, L6,
  \dodoi{10.3847/2041-8213/aa78fa}

\bibitem[{Spekkens {et~al.}(2014)Spekkens, Urbancic, Mason, Willman, \&
  Aguirre}]{spekkens_dearth_2014}
Spekkens, K., Urbancic, N., Mason, B.~S., Willman, B., \& Aguirre, J.~E. 2014,
  ApJ, 795, L5, \dodoi{10.1088/2041-8205/795/1/L5}

\bibitem[{{Springel}(2010)}]{springel_arepo_2010}
{Springel}, V. 2010, \mnras, 401, 791, \dodoi{10.1111/j.1365-2966.2009.15715.x}

\bibitem[{Stadel(2001)}]{stadel_cosmological_2001}
Stadel, J.~G. 2001, Ph.D. Thesis, 3657

\bibitem[{Stinson {et~al.}(2006)Stinson, Seth, Katz, Wadsley, Governato, \&
  Quinn}]{stinson_star_2006}
Stinson, G., Seth, A., Katz, N., {et~al.} 2006, MNRAS, 373, 1074,
  \dodoi{10.1111/j.1365-2966.2006.11097.x}

\bibitem[{Stinson {et~al.}(2012)Stinson, Brook, Prochaska, Hennawi, Shen,
  Wadsley, Pontzen, Couchman, Quinn, Macci{\`o}, \&
  Gibson}]{stinson_magicc_2012}
Stinson, G.~S., Brook, C., Prochaska, J.~X., {et~al.} 2012, Mon Not R Astron
  Soc, 425, 1270, \dodoi{10.1111/j.1365-2966.2012.21522.x}

\bibitem[{{Tonnesen} \& {Bryan}(2009)}]{Tonnesen2009}
{Tonnesen}, S., \& {Bryan}, G.~L. 2009, \apj, 694, 789,
  \dodoi{10.1088/0004-637X/694/2/789}

\bibitem[{Tremmel {et~al.}(2017)Tremmel, Karcher, Governato, Volonteri, Quinn,
  Pontzen, Anderson, \& Bellovary}]{tremmel_romulus_2017}
Tremmel, M., Karcher, M., Governato, F., {et~al.} 2017, MNRAS, 470, 1121,
  \dodoi{10.1093/mnras/stx1160}

\bibitem[{{Tremmel} {et~al.}(2019){Tremmel}, {Quinn}, {Ricarte}, {Babul},
  {Chadayammuri}, {Natarajan}, {Nagai}, {Pontzen}, \&
  {Volonteri}}]{Tremmel2019}
{Tremmel}, M., {Quinn}, T.~R., {Ricarte}, A., {et~al.} 2019, \mnras, 483, 3336,
  \dodoi{10.1093/mnras/sty3336}

\bibitem[{Trentham \& Tully(2009)}]{trentham_dwarf_2009}
Trentham, N., \& Tully, R.~B. 2009, MNRAS, 398, 722,
  \dodoi{10.1111/j.1365-2966.2009.15189.x}

\bibitem[{van~der Walt {et~al.}(2011)van~der Walt, Colbert, \&
  Varoquaux}]{walt_numpy_2011}
van~der Walt, S., Colbert, S.~C., \& Varoquaux, G. 2011, Computing in Science
  Engineering, 13, 22, \dodoi{10.1109/MCSE.2011.37}

\bibitem[{Wadsley {et~al.}(2017)Wadsley, Keller, \&
  Quinn}]{wadsley_gasoline2_2017}
Wadsley, J.~W., Keller, B.~W., \& Quinn, T.~R. 2017, \mnras, 471, 2357,
  \dodoi{10.1093/mnras/stx1643}

\bibitem[{Wadsley {et~al.}(2004)Wadsley, Stadel, \&
  Quinn}]{wadsley_gasoline_2004}
Wadsley, J.~W., Stadel, J., \& Quinn, T. 2004, New Astronomy, 9, 137,
  \dodoi{10.1016/j.newast.2003.08.004}

\bibitem[{{Weidemann}(1987)}]{Weidemann1987}
{Weidemann}, V. 1987, \aap, 188, 74

\bibitem[{Weisz {et~al.}(2014)Weisz, Dolphin, Skillman, Holtzman, Gilbert,
  Dalcanton, \& Williams}]{weisz_star_2014}
Weisz, D.~R., Dolphin, A.~E., Skillman, E.~D., {et~al.} 2014, ApJ, 789, 148,
  \dodoi{10.1088/0004-637X/789/2/148}

\bibitem[{Weisz {et~al.}(2015)Weisz, Dolphin, Skillman, Holtzman, Gilbert,
  Dalcanton, \& Williams}]{weisz_star_2015}
---. 2015, ApJ, 804, 136, \dodoi{10.1088/0004-637X/804/2/136}

\bibitem[{Wetzel {et~al.}(2015{\natexlab{a}})Wetzel, Deason, \&
  {Garrison-Kimmel}}]{wetzel_satellite_2015}
Wetzel, A.~R., Deason, A.~J., \& {Garrison-Kimmel}, S. 2015{\natexlab{a}}, ApJ,
  807, 49, \dodoi{10.1088/0004-637X/807/1/49}

\bibitem[{{Wetzel} {et~al.}(2016){Wetzel}, {Hopkins}, {Kim},
  {Faucher-Gigu{\`e}re}, {Kere{\v{s}}}, \& {Quataert}}]{Wetzel2016}
{Wetzel}, A.~R., {Hopkins}, P.~F., {Kim}, J.-h., {et~al.} 2016, \apjl, 827,
  L23, \dodoi{10.3847/2041-8205/827/2/L23}

\bibitem[{Wetzel {et~al.}(2012)Wetzel, Tinker, \& Conroy}]{wetzel_galaxy_2012}
Wetzel, A.~R., Tinker, J.~L., \& Conroy, C. 2012, \mnras, 424, 232,
  \dodoi{10.1111/j.1365-2966.2012.21188.x}

\bibitem[{Wetzel {et~al.}(2013)Wetzel, Tinker, Conroy, \& {van den
  Bosch}}]{wetzel_galaxy_2013}
Wetzel, A.~R., Tinker, J.~L., Conroy, C., \& {van den Bosch}, F.~C. 2013,
  MNRAS, 432, 336, \dodoi{10.1093/mnras/stt469}

\bibitem[{Wetzel {et~al.}(2015{\natexlab{b}})Wetzel, Tollerud, \&
  Weisz}]{wetzel_rapid_2015}
Wetzel, A.~R., Tollerud, E.~J., \& Weisz, D.~R. 2015{\natexlab{b}}, ApJ, 808,
  L27, \dodoi{10.1088/2041-8205/808/1/L27}

\bibitem[{Wheeler {et~al.}(2014)Wheeler, Phillips, Cooper, {Boylan-Kolchin}, \&
  Bullock}]{wheeler_surprising_2014}
Wheeler, C., Phillips, J.~I., Cooper, M.~C., {Boylan-Kolchin}, M., \& Bullock,
  J.~S. 2014, MNRAS, 442, 1396, \dodoi{10.1093/mnras/stu965}

\bibitem[{{Wilson}(1927)}]{Wilson1927}
{Wilson}, E.~B. 1927, Journal of the American Statistical Association, 22, 209,
  \dodoi{10.1080/01621459.1927.10502953}

\bibitem[{Wright {et~al.}(2019)Wright, Brooks, Weisz, \&
  Christensen}]{wright_reignition_2019}
Wright, A.~C., Brooks, A.~M., Weisz, D.~R., \& Christensen, C.~R. 2019, MNRAS,
  482, 1176, \dodoi{10.1093/mnras/sty2759}

\bibitem[{Zolotov {et~al.}(2012)Zolotov, Brooks, Willman, Governato, Pontzen,
  Christensen, Dekel, Quinn, Shen, \& Wadsley}]{zolotov_baryons_2012}
Zolotov, A., Brooks, A.~M., Willman, B., {et~al.} 2012, ApJ, 761, 71,
  \dodoi{10.1088/0004-637X/761/1/71}

\end{thebibliography}

\end{document}